\documentclass[aps, prd, amsmath, amssymb, amsfonts, floats, %
floatfix, superscriptaddress, nofootinbib, twocolumn, showpacs]%
{revtex4}

\allowdisplaybreaks[2]

\usepackage{graphicx}
\usepackage[usenames, dvipsnames]{color}
\usepackage[colorlinks, pdfborder={0 0 0}, plainpages=false]{hyperref}
\usepackage{breakurl}
\usepackage{amsmath, amssymb, amsfonts}
\usepackage{xspace} 
\usepackage{dcolumn}
\usepackage{bm}
\usepackage{multirow} 

\definecolor{CiteColor}{rgb}{0, 0.5, 0}
\hypersetup{citecolor=CiteColor} %
\definecolor{RefColor}{rgb}{0.55, 0, 0}
\hypersetup{linkcolor=RefColor} %

\usepackage{ulem}
\definecolor {darkgreen}{rgb}{0.2, 0.7, 0.2}

\newcommand{\np}{({\bf n}\cdot{\bf \hat{p}})}
\newcommand{\p}{\boldsymbol{\hat{p}}}

\newcommand{\Maryland}{\affiliation{Maryland Center for Fundamental
    Physics \& Joint Space-Science Institute \\
  Department of Physics, University of Maryland, College
    Park, MD 20742}}

\begin{document}

\title{Extending the effective-one-body Hamiltonian of black-hole binaries to include 
next-to-next-to-leading spin-orbit couplings}

\author{Enrico Barausse} \Maryland %
\author{Alessandra Buonanno} \Maryland %

\begin{abstract}
In the effective-one-body (EOB) approach the dynamics of two compact objects of masses $m_1$ and $m_2$
and spins $\mathbf{S}_1$ and $\mathbf{S}_2$ is mapped into the dynamics of one
test particle of mass $\mu = m_1\,m_2/(m_1+m_2)$ and spin $\mathbf{S}_*$ 
moving in a deformed Kerr metric with mass $M =m_1+m_2$ and spin $\mathbf{S}_\mathrm{Kerr}$. 
In a previous paper we computed an EOB Hamiltonian for spinning 
black-hole binaries that (i) when expanded in post-Newtonian orders, 
reproduces the leading order spin-spin coupling and the leading and next-to-leading order
spin-orbit couplings for any mass ratio, and (iii) reproduces
{\it all} spin-orbit couplings in the test-particle limit. Here we extend this EOB 
Hamiltonian to include next-to-next-to-leading spin-orbit couplings for 
any mass ratio. We discuss two classes of EOB Hamiltonians that differ by the way 
the spin variables are mapped between the effective and real descriptions. We also 
investigate the main features of the dynamics 
when the motion is equatorial, such as the existence of the 
innermost stable circular orbit and of a peak in the orbital 
frequency during the plunge subsequent to the inspiral. 
\end{abstract}

\date{\today \hspace{0.2truecm}}


\maketitle

\section{Introduction}
\label{sec:intro}

Coalescing compact binaries composed of neutron stars and/or black holes 
are among the most promising gravitational-wave sources for 
ground-based detectors, such as the Laser Interferometer
Gravitational-wave Observatory (LIGO)~\cite{Abbott:2007}, Virgo~\cite{Acernese:2008}, 
GEO~\cite{Grote:2008zz}, the Large Cryogenic Gravitational Telescope (LCGT)~\cite{Kuroda:2010}, 
and future space-based detectors. 

So far, the search for gravitational waves with LIGO, GEO and Virgo detectors has
focused on non-spinning compact binaries~\cite{Abbott:2005kq,Abbott:2007xi,
Abbott:2009qj,Abadie:2010yba,Abadie:2011kd}, although in Ref.~\cite{Abbott:2007ai} 
single-spin templates were used to search for inspiraling 
spinning compact objects. Within the next 4-5 years LIGO and Virgo detectors will be 
upgraded to a sensitivity such that event rates for coalescing binary systems will 
increase by a factor of one thousand. Thus, it is timely and necessary to develop more accurate templates 
that include spin effects. For maximally spinning objects, we expect that reasonably accurate 
templates would need to be computed at least through 3.5PN order. In the non-spinning case, studies 
at the interface between numerical and analytical relativity have demonstrated that templates 
computed at 3.5PN order are indeed reasonably accurate.

In the last few years, motivated by the search for gravitational waves, the knowledge 
of spin effects in the two-body dynamics and gravitational-wave emission 
within the post-Newtonian (PN)~\footnote{We refer to $n$PN as the order equivalent to terms $O(c^{-2n})$
in the equations of motion beyond the Newtonian acceleration.} approximation has improved considerably. 
In particular, spin-orbit (SO) effects in the two-body equations of motion 
are currently known through 3.5PN order (i.e., 2PN order beyond the 
leading SO term)~\cite{kidder93,OTO98,TOO01,Faye-Blanchet-Buonanno:2006,
DJSspin,Porto:2010tr,Perrodin:2010dy,Levi:2010zu,Hartung:2011te}, and in the 
energy flux through 3PN order~\cite{kidder95,OTO98,TOO01,W05,Blanchet-Buonanno-Faye:2006,Blanchet:2011zv} 
(i.e., 1.5PN order beyond the leading SO term). Moreover, spin-spin (SS) effects have been 
computed through 3PN order (i.e., 1PN order beyond the leading SS term) in 
the conservative dynamics~\cite{kidder93,kidder95,MVGer05,Porto06,Porto:2006bt,Hergt:2007ha,SHS07,
hergt_schafer_08,SHS08,SSH08,Porto:2008tb,PR08,Levi:2010} and also in the multipole 
moments~\cite{Porto:2010zg}.

In order to build reliable templates and search for gravitational-waves from high-mass 
compact binaries that merge in the detector bandwidth, it is crucial to improve the 
PN approximation by resumming the dynamics and gravitational emission in a suitable 
way and by using numerical relativity and perturbation theory as a guidance. 
The effective-one-body approach (EOB)~\cite{Buonanno00, Buonanno99, DJS00, Damour01c, Buonanno06} 
offers the possibility of fulfilling this goal. The EOB 
approach uses the results of PN theory, not in 
their original Taylor-expanded form (i.e., as
polynomials in $v/c$), but instead in a suitably resummed form.
In particular, it maps the dynamics of two compact objects of masses $m_1$ and $m_2$,
and spins $\mathbf{S}_1$ and $\mathbf{S}_2$, into the dynamics of one
test particle of mass $\mu = m_1\,m_2/(m_1+m_2)$ and spin
$\mathbf{S}_*$ moving in a deformed Kerr metric with mass $M =
m_1+m_2$ and spin $\mathbf{S}_\mathrm{Kerr}$. The deformation parameter
is the symmetric mass ratio $\eta=m_1\,m_2/(m_1+m_2)^2$, which ranges
between $0$ (test particle limit) and $1/4$ (equal-mass limit). 
The analyses and theoretical progress made in Refs.~\cite{Buonanno-Cook-Pretorius:2007,
  Buonanno2007, Pan2007, Boyle2008a, Buonanno:2009qa, Barausse:2009xi, Pan:2009wj, Pan2010hz,
  Damour2007a, DN2007b, DN2008, DIN, Damour2009a, Yunes:2009ef, Yunes:2010zj,Bernuzzi:2010xj,Pan:2011gk} have
demonstrated that faithful EOB templates describing the full signal (i.e., the inspiral, 
merger and ringdown) can be built and used in real searches~\cite{Abadie:2011kd}.

Here we build on previous work~\cite{Damour01c,Damour:2007nc,Barausse:2009aa,Barausse:2009xi}, 
employ the recent results of Ref.~\cite{Hartung:2011te} and extend the EOB conservative 
dynamics, i.e. the EOB Hamiltonian, through 3.5PN order in the SO couplings. Since the mapping 
between the PN-expanded Hamiltonian (or real Hamiltonian) and the EOB Hamiltonian is not unique, we explore 
two specific classes of EOB Hamiltonians, which differ by the way the spin variables of the real and effective 
descriptions are mapped.

This paper is organized as follows. In Sec.~\ref{sec:SEOBH}, after reviewing the 
logic underpinning the construction of the EOB Hamiltonian, we proceed in steps 
and extend the EOB Hamiltonian proposed in Ref.~\cite{Barausse:2009xi} through 
3.5PN order in the SO couplings. In particular, in Sec.~\ref{sec:ADMH} we derive 
the PN-expanded Arnowitt-Deser-Misner (ADM) Hamiltonian in the EOB canonical 
coordinates; then, after computing in Sec.~\ref{sec:EffH} the effective Hamiltonian corresponding 
to the canonically transformed PN-expanded ADM Hamiltonian, we compare it (in Sec.~\ref{sec:PNexp}) 
to the deformed-Kerr Hamiltonian for a spinning test-particle~\cite{Barausse:2009xi}, 
and work out (in Secs.~\ref{sec:SEOBgeo} and \ref{sec:SEOBngeo}) two classes of EOB Hamiltonians. 
In Sec.~\ref{sec:EOBdynamics} we study the dynamics of these Hamiltonians for equatorial orbits, and  
in Sec.~\ref{sec:concl} we summarize our main conclusions.

We use geometric units $G = c = 1$ throughout the paper,
except when performing PN expansions, where 
powers of the speed of light $c$ are restored and
play the role of book-keeping parameters. 

\section{The effective-one-body Hamiltonian for two spinning black holes}
\label{sec:SEOBH}

The main ingredient of the EOB approach is the {\it real} PN-expanded  ADM Hamiltonian (or {\it real} 
Hamiltonian) describing two black holes with masses $m_1, m_2$ and spins 
$\boldsymbol{S}_1$, $\boldsymbol{S}_2$. The real Hamiltonian is then 
canonically transformed and subsequently \textit{mapped} to an {\it effective} Hamiltonian $H_{\rm eff}$
describing a test-particle of mass $\mu=m_1\,m_2/(m_1+m_2)$ and suitable
spin $\boldsymbol{S}^\ast$, moving in a \textit{deformed} Kerr metric of mass $M = m_1 + m_2$ and 
suitable spin $\boldsymbol{S}_{\rm Kerr}$. The deformation is regulated by the binary's symmetric mass-ratio 
parameter, $\eta=\mu/M$, and therefore disappears in the test-particle limit $\eta \rightarrow 0$. 
The so-called improved real (or EOB) Hamiltonian reads
\begin{equation}
\label{hreal0}
H^{\rm improved}_\mathrm{real} = M\,\sqrt{1+2\eta\,\left(\frac{H_{\rm eff}}{\mu}-1\right)}\,.
\end{equation}
The computation of the EOB Hamiltonian consists of several steps. We briefly review these steps and the 
underlying logic that  we will follow in the next sections:
\begin{enumerate}
\item[(i)] We apply a canonical transformation to the PN-expanded ADM Hamiltonian using the Lie 
method, obtaining the PN-expanded Hamiltonian 
in EOB canonical coordinates (see Sec.~\ref{sec:ADMH});  
\item[(ii)]  We compute the effective Hamiltonian corresponding to the canonically 
transformed PN-expanded ADM Hamiltonian (see Sec.~\ref{sec:EffH});
\item[(iii)] We PN-expand the deformed-Kerr Hamiltonian for a spinning test-particle derived in 
Ref.~\cite{Barausse:2009xi} 
(see Sec.~\ref{sec:PNexp});
\item[(iv)] We compare (ii) and (iii), and  work out the mapping between the 
spin variables in the real and effective descriptions, and compute the improved EOB Hamiltonian (see Secs.~\ref{sec:SEOBgeo} and 
\ref{sec:SEOBngeo}). 
\end{enumerate}

\subsection{The ADM Hamiltonian canonically transformed to EOB coordinates}
\label{sec:ADMH}

Following Ref.~\cite{Barausse:2009xi}, we denote the ADM canonical variables in the 
binary's center-of-mass frame with $\boldsymbol{r}^\prime$ and $\boldsymbol{p}^\prime$, 
and we introduce the following spin variables:
\begin{eqnarray}
\boldsymbol{\sigma}&=&\boldsymbol{S}_1+\boldsymbol{S}_2\,, \label{sigma}\\
\boldsymbol{\sigma}^\ast&=&\boldsymbol{S}_1\,\frac{m_2}{m_1}+\boldsymbol{S}_2\,\frac{m_1}{m_2}\,.
\label{sigmastar}
\end{eqnarray}
Henceforth, to keep track of the PN orders,  
we rescale the spins variables as 
$\boldsymbol{{\sigma}}^\ast \rightarrow \boldsymbol{\sigma}^\ast \,c$ and 
$\boldsymbol{{\sigma}} \rightarrow \boldsymbol{\sigma} \,c$.

We use the spin-independent part of the ADM Hamiltonian through 3PN order~\cite{DJS00}, 
and we include SO effects through 2PN order beyond the leading-order effects (1.5PN), 
thus through 3.5PN order. In particular, the ADM SO Hamiltonian at 3.5PN order 
was computed recently in Ref.~\cite{Hartung:2011te} (the ADM SO Hamiltonian at 1.5PN 
was computed in Ref.~\cite{Damour-Schafer:1988}, and at 2.5PN in Ref.~\cite{Damour:2007nc}). 
In the binary's center-of-mass, the ADM SO Hamiltonian reads
\begin{equation}
\label{HSO_adm}
 H^{\rm ADM}_{\rm SO}(\boldsymbol{r}^{\prime},\boldsymbol{p}^{\prime},\boldsymbol{\sigma}^\ast,\boldsymbol{\sigma}) = 
\frac{1}{c^3}\,\frac{\boldsymbol{L}^\prime}{r^{\prime\,3}}\cdot (g^{\rm ADM}_\sigma \,
\boldsymbol{\sigma}+ g^{\rm ADM}_{\sigma^\ast}\,\boldsymbol{\sigma}^\ast)\,, 
\end{equation}
where we indicate $\boldsymbol{L}^\prime = \boldsymbol{r}^\prime \times \boldsymbol{p}^\prime$ and 
\begin{subequations}
\begin{eqnarray}
g^{\rm ADM}_{\sigma} &=& 2 + \frac{1}{c^2}\, \left [\frac{19}{8}\, \eta\,\boldsymbol{\hat{p}}^{\prime \,2} + 
\frac{3}{2} \eta\, (\boldsymbol{n}^{\prime}\cdot \boldsymbol{\hat{p}}^\prime)^2 \right . \nonumber \\
&& \left. - (6 + 2\eta)\, \frac{M}{r^{\prime}} \right ] \nonumber \\
&& + \frac{1}{c^4}\, \left [ \frac{15}{16}\eta^2\,(\boldsymbol{n}^{\prime} \cdot \boldsymbol{\hat{p}}^{\prime})^4 
+  \frac{21}{2}(1+\eta)\,\left(\frac{M}{r^{\prime}}\right)^2 \right. \nonumber \\ 
&& \left. + \frac{1}{8}\eta\,(-9+22\eta)\,\boldsymbol{\hat{p}}^{\prime \,4} 
-\frac{1}{16}\eta\,(314+39\eta)\,\frac{M}{r^{\prime}}\,
\boldsymbol{\hat{p}}^{\prime \,2} \right . \nonumber \\
&& \left. -\frac{1}{16}\eta\,(256+45\eta)\,\frac{M}{r^{\prime}}\,
(\boldsymbol{n}^{\prime} \cdot \boldsymbol{\hat{p}}^{\prime})^2\right . \nonumber \\
&& \left. + \frac{3}{16}\eta\,(-4+9\eta)\,\boldsymbol{\hat{p}}^{\prime \,2} \,
(\boldsymbol{n}^{\prime} \cdot \boldsymbol{\hat{p}}^{\prime})^2 \right ]\,, \\
g^{\rm ADM}_{\sigma^\ast} &=& \frac{3}{2} + \frac{1}{c^2}\, \left [ 
\left (-\frac{5}{8} + 2 \eta \right )\,\boldsymbol{\hat{p}}^{\prime\,2} + \frac{3}{4} \eta \,
(\boldsymbol{n}^{\prime} \cdot \boldsymbol{\hat{p}}^\prime)^2 \right . \nonumber \\
&& \left . - (5 + 2 \eta) \frac{M}{r^{\prime}} \right ] \nonumber \\
&& + \frac{1}{c^4}\,\left [\frac{1}{8}(75+82 \eta)\,\left(\frac{M}{r^{\prime}}\right)^2 \right. \nonumber \\ 
&&\left.  + \frac{1}{16}(7-37\eta+39\eta^2)\,\boldsymbol{\hat{p}}^{\prime \,4} 
\right. \nonumber \\
&& \left. -\frac{3}{16}(-18+86\eta+13 \eta^2)\,\frac{M}{r^{\prime}}\,
\boldsymbol{\hat{p}}^{\prime \,2} \right . \nonumber \\
&& \left. -\frac{3}{16}\eta\,(32+15\eta)\,\frac{M}{r^{\prime}}\,
(\boldsymbol{n}^{\prime} \cdot \boldsymbol{\hat{p}}^{\prime})^2\right . \nonumber \\
&& \left. + \frac{9}{16}\eta\,(-1+2\eta)\,\boldsymbol{\hat{p}}^{\prime \,2} \,
(\boldsymbol{n}^{\prime} \cdot \boldsymbol{\hat{p}}^{\prime})^2 \right ]\,, 
\end{eqnarray}
\end{subequations}
with $\boldsymbol{n}^\prime= \boldsymbol{r}^\prime/r^\prime$, and
where we have introduced the rescaled conjugate momentum
$\boldsymbol{\hat{p}}^\prime = \boldsymbol{p}^\prime/\mu$.

In order to canonically transform the ADM Hamiltonian to EOB
coordinates, various approaches are possible. A popular method, used
in the previous work on the EOB model~\cite{Buonanno99,DJS00,Damour:2007nc}, is to use 
a generating function that produces a near-identity transformation, i.e. one of the form
$\tilde{G}(q',\pi)=q^{\prime i} \pi_i+\epsilon\,{G}(q',\pi)$, where
$(q,\pi)$ are the phase variables (including the angles defining the
spins and their conjugate momenta, see Ref.~\cite{Barausse:2009aa}) and $\epsilon$ is a small
parameter.  Expressing the initial ``primed'' coordinates (the ADM
coordinates) in terms of the new ``unprimed'' coordinates (the EOB
coordinates), one gets
\begin{subequations}
\label{qtransf}
\begin{align}
q^{\prime i}=&q^i-\epsilon\frac{\partial G(q',\pi)}{\partial \pi_i}=q^{i}-\epsilon\,\frac{\partial G(q,\pi)}{\partial \pi_i}
\nonumber\\&+\epsilon^2\,\frac{\partial^2 G(q,\pi)}{\partial \pi_i\partial q^{j}}\,
\frac{\partial G(q,\pi)}{\partial \pi_{j}}
+{\cal O} (\epsilon^3)
\,,\\
\pi^{\prime}_i=&\pi_{i}+\epsilon\,\frac{\partial G(q',\pi)}{\partial q^{\prime i}}=
\pi_{i}+\epsilon\,\frac{\partial G(q,\pi)}{\partial q^{i}}
\nonumber\\&-\epsilon^2\,\frac{\partial^2 G(q,\pi)}{\partial q^{i}\partial q^{j}}\,\frac{\partial G(q,\pi)}{\partial \pi_{j}}
+{\cal O} (\epsilon^3)\,.
\end{align}
\end{subequations}
Because under a time-independent canonical transformation the Hamiltonian transforms 
as $H(q,p)=H'(q',p')$, Eqs.~(\ref{qtransf}) imply
\begin{eqnarray}
\label{Htransf}
H(q,\pi)&=&H'(q',\pi')=H'(q,\pi)\nonumber \\
&& \!\!\!\!\!\!\!\! +\epsilon\,\left[ 
\frac{\partial H'(q,\pi)}{\partial \pi_{i}}\,\frac{\partial G(q,\pi)}{\partial q^{ i}} 
-\frac{\partial H'(q,\pi)}{\partial q^{ i}}\,\frac{\partial G(q,\pi)}{\partial \pi_{i}} 
\right] \nonumber \\
&& \!\!\!\!\!\!\!\! +\epsilon^2\,\left[  \frac{\partial H'(q,\pi)}{\partial q^{ i}}\,
\frac{\partial^2 G(q,\pi)}{\partial \pi_i\partial q^{j}}\,
\frac{\partial G(q,\pi)}{\partial \pi_{j}} \right. \nonumber \\
&&\!\!\!\!\!\!\!\! \left. -\frac{\partial H'(q,\pi)}{\partial \pi_{i}}\,\frac{\partial^2 G(q,\pi)}{\partial q^{i}\partial q^{j}}\,
\frac{\partial G(q,\pi)}{\partial \pi_{j}}
 \right. \nonumber \\
&&\!\!\!\!\!\!\!\!  \left.
+\frac12  \frac{\partial^2 H'(q,\pi)}{\partial q^{ i}\partial q^{ j}}\,\frac{\partial G(q,\pi)}{\partial \pi_{i}}\,
\frac{\partial G(q,\pi)}{\partial \pi_{j}} 
\right. \nonumber \\
&& \!\!\!\!\!\!\!\! \left.+\frac12  \frac{\partial^2 H'(q,\pi)}{\partial \pi_{i}\partial \pi_{j}}\,
\frac{\partial G(q,\pi)}{\partial q^{ i}}\, \frac{\partial G(q,\pi)}{\partial q^{ j}}
\right. \nonumber \\
&& \!\!\!\!\!\!\!\! \left.- \frac{\partial^2 H'(q,\pi)}{\partial q^{ i}\partial \pi_{j}}\,
\frac{\partial G(q,\pi)}{\partial \pi_{i}}\,\frac{\partial G(q,\pi)}{\partial q^{ j}} 
\right] +{\cal O} (\epsilon^3)\,.
\end{eqnarray}
The terms of order ${\cal O}(\epsilon)$ in this equation can be
rewritten as $\epsilon\,\{G,H'\}$, which is very convenient because it
transforms a sum over all the phase variables (including the angles
defining the spins and their conjugate momenta) into a Poisson bracket
that can be computed using only the commutation relations
$\{x^i,p_j\}=\delta^i_j$, $\{x^i,S_{(a)}^j\}=0$,
$\{p_i,S_{(a)}^j\}=0$, and $\{S_{(a)}^i,
S_{(b)}^j\}=\delta_{(a)(b)}\epsilon_{ijk} S_{(a)}^k$ ($a,b=1,2$ being
indices that distinguish between the two black holes).  Unfortunately, the
terms ${\cal O}(\epsilon^2)$ cannot be easily expressed in terms of
Poisson brackets, which makes them hard to compute (because the spin
variables must be carefully taken into account in the sums). Also, the
generalization of Eq.~\eqref{Htransf} to higher orders in $\epsilon$ becomes 
more and more complicated.

A possible alternative to the generating function method mentioned above is
given by the so-called Lie method~\cite{Benettin:2004}. This approach exploits the fact
that the flux of the Hamilton equations is canonical. Therefore, one
can define a fictitious Hamiltonian ${\cal H}(q,\pi)$ whose flux 
sends some initial data $(q,\pi)$ to $(q'(q,\pi,\epsilon),\pi'(q,\pi,\epsilon))$, 
where $\epsilon$ is the ``time'' variable of this fictitious Hamiltonian. 
The canonical transformation is then simply given by $(q'(q,\pi,\epsilon),\pi'(q,\pi,\epsilon))$. 
The advantage of this approach is that any function $f(q,\pi)$ satisfies $\dot{f}=\{f,{\cal H}\}$ 
(where we denote with  $\dot{\phantom{a}}=d/d\epsilon$) under the Hamiltonian flux of ${\cal H}$. 
Defining for convenience ${\cal G}=-{\cal H}$,
this equation becomes $\dot{f}=\{{\cal G},f\}$, and denoting the
differential operator $\{{\cal G},\ldots\}$ by ${\cal L}_{{\cal G}}$,
a Taylor expansion yields
\begin{eqnarray}
f(q'(q,\pi,\epsilon),\pi'(q,\pi,\epsilon)) &=&\sum_{n=0}^{\infty}\frac{\epsilon^n}{n!}\,{\cal L}^n_{{\cal G}}\,f(q,\pi)\nonumber \\
&=&\exp{(\epsilon\,{\cal L}_{{\cal G}})}\,f(q,\pi) \nonumber \\
&=& f(q,\pi)+\epsilon\,\{{\cal G},f\}(q,\pi) \nonumber \\
&+&\frac12 \epsilon^2\,\{{\cal G},\{{\cal G},f\}\}(q,\pi)+{\cal O} (\epsilon^3)\,. \nonumber \\
\end{eqnarray}
Specializing to the (non-fictitious) Hamiltonian $H'$, we obtain the equivalent of Eq.~\eqref{Htransf}, that is
\begin{eqnarray}
\label{HtransfLie}
H(q,\pi)&=&H'(q',\pi')=H'(q,\pi)+\epsilon\,\{{\cal G},H'\}(q,\pi) \nonumber \\
&+& \frac12 \epsilon^2\, \{ {\cal G},\{ {\cal G},H' \}  \}(q,\pi) +{ \cal O } (\epsilon^3) \,.
\end{eqnarray}
As already mentioned, the above expression allows us to account for the
spin variables very easily, if necessary~\footnote{The Poisson brackets
  of the spin variables with themselves do not enter in the
  computations that we perform in this paper, but they do enter at
  higher PN orders.}, by means of the commutation relations
$\{x^i,S_{(a)}^j\}=0$, $\{p_i,S_{(a)}^j\}=0$, and $\{S_{(a)}^i,
S_{(b)}^j\}=\delta_{(a)(b)}\epsilon_{ijk} S_{(a)}^k$.

In this paper we will use the Lie method to generate the canonical transformation from ADM 
to EOB coordinates. In particular, we assume
\begin{equation}
{\cal G}(\boldsymbol{r},\boldsymbol{p})=\,\boldsymbol{r}\cdot
\boldsymbol{p}+{\cal G}_{\rm NS}(\boldsymbol{r},\boldsymbol{p})
+{\cal G}_{\rm S}(\boldsymbol{r},\boldsymbol{p},\boldsymbol{\sigma}^\ast,\boldsymbol{\sigma})\,,
\end{equation}
where ${\cal G}_{\rm NS}$ is the purely orbital part of the fictitious Hamiltonian, while
${\cal G}_{\rm S}$ is the spin-dependent part, which we assume to be linear in the spins since in this paper
we focus on the SO terms only. Because the transformations \eqref{Htransf} and \eqref{HtransfLie} agree 
at leading order in the perturbative parameter $\epsilon$, $G$ and ${\cal G}$ must agree at leading PN order. 
In particular, since the purely orbital generating function for the transformation from ADM to EOB coordinates
starts at 1PN, ${\cal G}_{\rm NS}$ must start at 1PN order too, that is 
\begin{multline}
{\cal G}_{\rm NS}(\boldsymbol{r},\boldsymbol{p})={\cal G}_{\rm NS\, 1PN}(\boldsymbol{r},\boldsymbol{p})
 +{\cal G}_{\rm NS\, 2PN}(\boldsymbol{r},\boldsymbol{p})
 +{\cal O}\left(\frac{1}{c^6}\right)\,,
\end{multline}
where ${\cal G}_{\rm NS\, 1PN}$ must coincide with $G_{\rm NS\, 1PN}$, and therefore be given by~\cite{Buonanno99}
\begin{equation}
{\cal G}_{\rm NS\, 1PN}(\boldsymbol{r},\boldsymbol{p})=\frac{1}{c^2}\, 
(\boldsymbol{r}\cdot\boldsymbol{p})\,\left[-\frac12 \eta\,\boldsymbol{{\hat{p}}}^{2} 
+ \frac{M}{r}\left(1+\frac12\eta\right)\,\right]\,.
\end{equation}
At 2PN, instead, ${\cal G}_{\rm NS}$ does not coincide with ${G}_{\rm NS}$, but a computation similar to the 
one in Ref.~\cite{Buonanno99} easily shows that 
\begin{eqnarray}
{\cal G}_{\rm NS\, 2PN}(\boldsymbol{r},\boldsymbol{p}) &=& \frac{1}{c^4} 
\,(\boldsymbol{r} \cdot \boldsymbol{p})\,\left [ \alpha\,\boldsymbol{\hat{p}}^4 + 
\beta\,\frac{M}{r}\,\boldsymbol{\hat{p}}^2 \right. \nonumber \\
&&\left.  + \gamma\,\frac{M}{r}\,(\boldsymbol{n} \cdot \boldsymbol{\hat{p}})^2 + \delta\,\left (\frac{M}{r}\right )^2 \right ]\,, 
\label{g2pn}
\end{eqnarray}
with
\begin{subequations}
\begin{eqnarray}
&&\alpha = \frac{\eta}{8} \,, \quad \quad \quad \quad \beta = \frac{\eta}{4}\,(4 - \eta)\,, \\
&&\gamma = \eta\,\frac{4+\eta}{8} \,, \quad \quad  \delta = \frac{1-7\eta + \eta^2}{4} \,.
\end{eqnarray}
\end{subequations}
[Note that the functional form \eqref{g2pn} is the same as for $G_{\rm NS\, 2PN}$, 
but the values of the parameters $\alpha$, $\beta$, $\gamma$
and $\delta$ are different from those of Ref.~\cite{Buonanno99}.]

Similarly, the spin-dependent part of the fictitious Hamiltonian, ${\cal G}_{\rm S}$, must start like $G_{\rm S}$ at 2.5PN order:
\begin{align}
 {\cal G}_{\rm S}(\boldsymbol{r},\boldsymbol{p},\boldsymbol{\sigma}^\ast,\boldsymbol{\sigma})&={\cal G}_{\rm S\, 2.5PN}(\boldsymbol{r},\boldsymbol{p},
 \boldsymbol{\sigma}^\ast,\boldsymbol{\sigma})  \nonumber\\&+
 {\cal G}_{\rm S\, 3.5PN}(\boldsymbol{r},\boldsymbol{p},\boldsymbol{\sigma}^\ast,\boldsymbol{\sigma})+{\cal O}\left(\frac{1}{c^9}\right)\,.
\end{align}
and if we restrict to functions that are linear in the spin variables, it must be~\cite{DJSspin}
\begin{align}
&{\cal G}_{\rm S\, 2.5PN}(\boldsymbol{r},\boldsymbol{p},\boldsymbol{\sigma}^\ast,\boldsymbol{\sigma})= \nonumber \\
& \quad
\frac{1}{c^5\,r^3}\, (\boldsymbol{r} \cdot\boldsymbol{\hat{p}})\,\left[a_0(\eta)\, (\boldsymbol{L} \cdot \boldsymbol{\sigma}) 
 + b_0(\eta)\, (\boldsymbol{L} \cdot\boldsymbol{\sigma}^\ast)\right]\,,\label{g25pnForm}
\end{align}
where $a_0(\eta)$ and $b_0(\eta)$ are arbitrary gauge functions. [Note that restricting to functions that are linear in the spin 
variables is justified because here we are looking at SO effects only, but in general cubic terms in the spin may be present, 
see Ref.~\cite{Barausse:2009xi}.]

The most general form for ${\cal G}_{\rm S}$ at 3.5PN order is instead, if we restrict again to functions
linear in the spins,
\begin{eqnarray}\label{g35pn}
{\cal G}_{\rm S\, 3.5 PN}(\boldsymbol{r},\boldsymbol{p}) &=&\frac{1}{c^7\,r^3}\,
(\boldsymbol{r}\cdot \boldsymbol{\hat{p}}) \left\{\left(\boldsymbol{L}\cdot\boldsymbol{\sigma}\right) \left[a_1(\eta) \,\boldsymbol
{\hat{p}}^2 \right. \right. \nonumber \\
&& \left. \left. + a_2(\eta) \,\frac{M}{r} + a_3(\eta) \,\left(\boldsymbol{n} \cdot \boldsymbol{\hat{p}}\right)^2\right] \right . 
\nonumber \\
&& \left . + \left(\boldsymbol{L}\cdot\boldsymbol{\sigma^\ast}\right) \left[b_1(\eta)\,\boldsymbol{\hat{p}}^2  
+ b_2(\eta)\,\frac{M}{r} \right. \right. \nonumber \\
&& \left . \left. + b_3(\eta) \left(\boldsymbol{n} \cdot \boldsymbol{\hat{p}}\right)^2\right]\right\}\,,
\end{eqnarray}
where $a_i(\eta)$ and $b_i(\eta)$ with $i=1,2,3$ are other arbitrary gauge functions. 
To ease the notation, henceforth we drop the $\eta$ dependence in the gauge parameters, both at 2.5PN and 3.5PN order, 
and will denote them simply with $a_i$ and $b_i$ (with $i=0,3$). 

Applying Eq.~\eqref{HtransfLie}, we obtain that the 3.5 SO Hamiltonian in EOB coordinates is given by
\begin{eqnarray}
\label{HADM}
H_{\rm SO \,3.5 PN} &=& H^{\rm ADM}_{\rm SO \,3.5 PN} 
+\{{\cal G}_{\rm NS\,2 PN}, H^{\rm ADM}_{\rm SO\, 1.5 PN}\}\nonumber \\
&& +\{{\cal G}_{\rm NS\, 1 PN}, H^{\rm ADM}_{\rm SO \,2.5 PN}\} \nonumber \\
&& + \{{\cal G}_{\rm S\, 2.5 PN}, H^{\rm ADM}_{\rm 1PN}\}  \nonumber \\
&& + \{{\cal G}_{\rm S\, 3.5 PN}, H^{\rm ADM}_{\rm Newt}\} \nonumber \\
&& +\frac12\{{\cal G}_{\rm NS\, 1 PN},\{{\cal G}_{\rm NS\, 1 PN}, H^{\rm ADM}_{\rm SO \,1.5 PN}\}\} \nonumber\\
&& +\frac12\{{\cal G}_{\rm NS\, 1 PN},\{{\cal G}_{\rm S\, 2.5 PN}, H^{\rm ADM}_{\rm Newt}\}\} \nonumber\\
&& +\frac12\{{\cal G}_{\rm S\, 2.5 PN},\{{\cal G}_{\rm NS\, 1 PN}, H^{\rm ADM}_{\rm Newt}\}\} \,. 
\nonumber \\
\end{eqnarray}
A tedious but straightforward calculation gives the several terms entering the above equation:
\begin{align}
\{{\cal G}_{\rm NS\, 2 PN},&H^{\rm ADM}_{\rm SO\, 1.5 PN}\} = \frac{3}{c^7\,r^3}\boldsymbol{L}\cdot\left(2\boldsymbol{\sigma}+
\frac32\boldsymbol{\sigma}^\ast\right)\,\nonumber \\ 
& \left[ \delta\,\left(\frac{M}{r}\right)^2 + \alpha\,\boldsymbol{\hat{p}}^{4} 
+ \beta\,\frac{M}{r}\,\boldsymbol{\hat{p}}^{2} + 4 \alpha\,\boldsymbol{\hat{p}}^{2} \,
(\boldsymbol{n} \cdot \boldsymbol{\hat{p}})^2 \right. \nonumber \\
& \left. + (2\beta +3\gamma)\,\frac{M}{r}\,(\boldsymbol{n} \cdot \boldsymbol{\hat{p}})^2 
\right ]\,,
\end{align}
\begin{align}
\{{\cal G}_{\rm NS\, 1 PN},& H^{\rm ADM}_{\rm SO \,2.5 PN}\} = \nonumber \\
& \frac{1}{c^7\,r^3}\boldsymbol{L}\cdot\boldsymbol{\sigma}\,
\left[ -4(6+5\eta+\eta^2)\,\left(\frac{M}{r}\right)^2 \right.\nonumber\\&\left.-\frac{95}{16}\eta^2\,\boldsymbol{\hat{p}}^{4} 
+ \frac{1}{16}\eta\,(382+159\eta)\,\frac{M}{r}\,\boldsymbol{\hat{p}}^{2} 
\right.\nonumber\\&\left.-\frac{63}{8}\eta^2\, \boldsymbol{\hat{p}}^{2} \,(\boldsymbol{n} \cdot \boldsymbol{\hat{p}})^2  
-\frac{15}{2}\eta^2\, (\boldsymbol{n} \cdot \boldsymbol{\hat{p}})^4\right.\nonumber\\&\left. + \frac{1}{8}\eta\,(190+63 \eta)\,\frac{M}{r}\,(\boldsymbol{n} \cdot \boldsymbol{\hat{p}})^2 \right ] 
\nonumber\\&+ \frac{1}{c^7\,r^3}\boldsymbol{L}\cdot\boldsymbol{\sigma}^\ast\, 
\left[ -2(10 + 9 \eta + 2\eta^2)\,\left(\frac{M}{r}\right)^2 \right. \nonumber \\
& \left. + \frac{5}{16}\eta (5 - 16 \eta)\,\boldsymbol{\hat{p}}^{4} 
+ \frac{1}{16}(-50 + 295\eta + 144 \eta^2)\,\times \right. \nonumber \\
& \left. \frac{M}{r}\,\boldsymbol{\hat{p}}^{2} + \frac{3}{8}\eta\,(5 -17 \eta)\,\boldsymbol{\hat{p}}^{2} \,(\boldsymbol{n} \cdot \boldsymbol{\hat{p}})^2 
 -\frac{15}{4}\eta^2\,(\boldsymbol{n} \cdot \boldsymbol{\hat{p}})^4 \right. \nonumber \\ 
& \left .+ \frac{1}{8}(10 + 151\eta+57 \eta^2)\,\frac{M}{r}\,(\boldsymbol{n} \cdot \boldsymbol{\hat{p}})^2 
\right ]\,,
\end{align}
\begin{align}
&  \{{\cal G}_{\rm S\, 2.5 PN}, H^{\rm ADM}_{\rm 1PN}\} = \frac{1}{c^7\,r^3}
  (a_0\, \boldsymbol{L}\cdot\boldsymbol{\sigma} +  b_0\,\boldsymbol{L}\cdot\boldsymbol{\sigma^\ast} )\times
\nonumber \\
&\quad \left [\left(\frac{M}{r}\right)^2 
+ \frac{1}{2}(-1 + 3\eta)\,\boldsymbol{\hat{p}}^{4}  -\frac{3}{2}(3+\eta)\,\frac{M}{r}\,\boldsymbol{\hat{p}}^{2} \right. \nonumber \\
&\quad \left. + \frac{9}{2}(2+\eta)\, \frac{M}{r}\,(\boldsymbol{n} \cdot \boldsymbol{\hat{p}})^2 -\frac{3}{2}(-1+3\eta)\,\boldsymbol{\hat{p}}^{2} \,
(\boldsymbol{n} \cdot \boldsymbol{\hat{p}})^2 \right ]\,,\nonumber\\
\end{align}
\begin{widetext}
\begin{eqnarray}
  \{{\cal G}_{\rm S\, 3.5 PN}, H^{\rm ADM}_{\rm Newt}\} &=& \frac{1}{c^7\,r^3}
  \boldsymbol{L}\cdot\boldsymbol{\sigma}\,\left [ -a_2\,\left(\frac{M}{r}\right)^2 +a_1\,\boldsymbol{\hat{p}}^{4} 
+(a_2-a_1)\,\frac{M}{r}\,\boldsymbol{\hat{p}}^{2} -(4a_2+2a_1+3a_3)\, \frac{M}{r}\,
(\boldsymbol{n} \cdot \boldsymbol{\hat{p}})^2 \right .\nonumber \\
&& \left. -3(a_1-a_3)\,\boldsymbol{\hat{p}}^{2} \,
(\boldsymbol{n} \cdot \boldsymbol{\hat{p}})^2 -5a_3\, (\boldsymbol{n} \cdot \boldsymbol{\hat{p}})^4 \right ] + \frac{1}{c^7\,r^3}
  \boldsymbol{L}\cdot\boldsymbol{\sigma^\ast}\,\left [ -b_2\,\left(\frac{M}{r}\right)^2 +b_1\,\boldsymbol{\hat{p}}^{4} +(b_2-b_1)\,\frac{M}{r}\,\boldsymbol{\hat{p}}^{2} \right. \nonumber \\
&& \left. -(4b_2+2b_1+3b_3)\, \frac{M}{r}\,
(\boldsymbol{n} \cdot \boldsymbol{\hat{p}})^2 -3(b_1-b_3)\,\boldsymbol{\hat{p}}^{2} \,
(\boldsymbol{n} \cdot \boldsymbol{\hat{p}})^2 -5b_3\,(\boldsymbol{n} \cdot \boldsymbol{\hat{p}})^4 \right ]\,,
\end{eqnarray}
\begin{eqnarray}
\{{\cal G}_{\rm NS\, 1 PN},\{{\cal G}_{\rm NS\, 1 PN}, H^{\rm ADM}_{\rm SO \,1.5 PN }\}\} &=&
\frac{3}{4c^7\, r^3}\boldsymbol{L}\cdot\left(2\boldsymbol{\sigma}+\frac32\boldsymbol{\sigma}^\ast\right)\, 
 \left[ 4(2+\eta)^2\,\left(\frac{M}{r}\right)^2 + 5\eta^2\,\boldsymbol{\hat{p}}^{4} 
-9\eta(2+\eta)\,\frac{M}{r}\,\boldsymbol{\hat{p}}^{2} \right. \nonumber \\
&& \left. + 8\eta^2\,\boldsymbol{\hat{p}}^{2} \,(\boldsymbol{n} \cdot \boldsymbol{\hat{p}})^2 
-12\eta\,(2+\eta)\, \frac{M}{r}\,(\boldsymbol{n} \cdot \boldsymbol{\hat{p}})^2 
+ 20\eta^2\,(\boldsymbol{n} \cdot \boldsymbol{\hat{p}})^4 \right ]\,,
\end{eqnarray}
\begin{eqnarray}
\{{\cal G}_{\rm NS\, 1 PN},\{{\cal G}_{\rm S\, 2.5 PN}, H^{\rm ADM}_{\rm Newt}\}\} &=&
\frac{1}{c^7\, r^3}\boldsymbol{L}\cdot\left(a_0\,\boldsymbol{\sigma}+b_0\,\boldsymbol{\sigma}^\ast\right)\,
\left[ -2(2+\eta)\left(\frac{M}{r}\right)^2 - \frac{5}{2}\eta\,\boldsymbol{\hat{p}}^{4} 
+\frac{1}{2}(10+9\eta)\,\frac{M}{r}\,\boldsymbol{\hat{p}}^{2} \right. \nonumber \\
&& \left. -\frac{3}{2}\eta\,\boldsymbol{\hat{p}}^{2} \,(\boldsymbol{n} \cdot \boldsymbol{\hat{p}})^2 
-\frac{1}{2}\,(22+3\eta)\, \frac{M}{r}\,(\boldsymbol{n} \cdot \boldsymbol{\hat{p}})^2 
+ 15\eta\,(\boldsymbol{n} \cdot \boldsymbol{\hat{p}})^4 \right ]\,,
\end{eqnarray}
\begin{eqnarray}
\{{\cal G}_{\rm S\, 2.5 PN},\{{\cal G}_{\rm NS\, 1 PN}, H^{\rm ADM}_{\rm Newt}\}\} &=&
\frac{1}{c^7\, r^3}\boldsymbol{L}\cdot\left(a_0\,\boldsymbol{\sigma}+b_0\,\boldsymbol{\sigma}^\ast\right)\,
\left[ -(2+\eta)\left(\frac{M}{r}\right)^2 - 2\eta\,\boldsymbol{\hat{p}}^{4} 
+3(1+\eta)\,\frac{M}{r}\,\boldsymbol{\hat{p}}^{2} \right. \nonumber \\
&& \left. +6\eta\,\boldsymbol{\hat{p}}^{2} \,(\boldsymbol{n} \cdot \boldsymbol{\hat{p}})^2 
-\frac{3}{2}\,(2+5\eta)\, \frac{M}{r}\,(\boldsymbol{n} \cdot \boldsymbol{\hat{p}})^2 \right ]\,.
\end{eqnarray}
\end{widetext}
Also, we have~\cite{Barausse:2009xi}
\begin{subequations}
\label{HADMothers}
\begin{eqnarray}
H_{\rm Newt} &=& H^{\rm ADM}_{\rm Newt}\,,\\
H_{\rm 1 PN} &=& H^{\rm ADM}_{\rm 1 PN} 
+\{{\cal G}_{\rm NS\,1 PN}, H^{\rm ADM}_{\rm Newt}\}\,,\label{h1transf}\\
H_{\rm SO \,1.5 PN} &=& H^{\rm ADM}_{\rm SO \,1.5 PN} \,,\\
H_{\rm SO \,2.5 PN} &=& H^{\rm ADM}_{\rm SO \,2.5 PN} 
+\{{\cal G}_{\rm S\,2.5 PN}, H^{\rm ADM}_{\rm Newt}\} \nonumber \\
&& + \{{\cal G}_{\rm 1 PN}, H^{\rm ADM}_{\rm SO\,1.5PN}\}\,,\label{h25transf}
\end{eqnarray}
\end{subequations}
where $H^{\rm ADM}_{\rm Newt}$, $H^{\rm ADM}_{\rm 1PN}$ can be found in Ref.~\cite{Buonanno99}, 
the explicit expressions of $\{{\cal G}_{\rm S\,2.5 PN}, H^{\rm ADM}_{\rm Newt}\}$ 
and $\{{\cal G}_{\rm 1 PN}, H^{\rm ADM}_{\rm SO\,1.5PN}\}$ 
are given in Eqs.~(5.20), (5.24) of Ref.~\cite{Barausse:2009xi}, while 
\begin{align}
\{{\cal G}_{\rm NS\,1 PN},&H^{\rm ADM}_{\rm Newt}\} = \frac{\mu}{c^2}\left [-\frac{1}{2}(2+\eta)\,
\left (\frac{M}{r}\right )^2 -\frac{\eta}{2}\,\boldsymbol{\hat{p}}^{4} \right. \nonumber \\
& \left. + (1+\eta)\,\frac{M}{r}\,\boldsymbol{\hat{p}}^{2} + \frac{1}{2}(-2+\eta)\,\frac{M}{r}\,(\boldsymbol{n}\cdot \boldsymbol{\hat{p}})^2
\right ]\,.
\end{align}
Also, we note that Eqs.~\eqref{h1transf} and \eqref{h25transf} immediately 
imply ${\cal G}_{\rm 2.5 PN}=G_{\rm 2.5PN}$, 
i.e. the 2.5 PN gauge parameters $a_0$ and $b_0$ appearing in Eq.~\eqref{g25pnForm} have the same meaning 
in the Lie method and in the generating function approaches.
\subsection{The spin-orbit terms in the effective Hamiltonian through 3.5PN order}
\label{sec:EffH}

Following Refs.~\cite{Buonanno99,2000PhRvD..62h4011D,Damour01c}, we map the effective and real two-body Hamiltonians as
\begin{equation}
\label{heff}
\frac{H_{\rm eff}}{\mu c^2} =
\frac{H_{\rm real}^2 - m_1^2\,c^4 - m_2^2\, c^4}
{2 m_1\, m_2\, c^4}\,,
\end{equation}
where $H_{\rm real}$ is the real two-body Hamiltonian containing also the rest-mass contribution $M\,c^2$. 
Expanding Eq.~(\ref{heff}) in powers of $1/c$, we have
\begin{eqnarray}
H^{\rm eff}_{\rm SO \,3.5PN} &=& H_{\rm SO \,3.5 PN} +\frac{1}{M}\,(H_{\rm SO\,1.5 PN}\,H_{\rm 1 PN} \nonumber \\
&& + H_{\rm SO\,2.5 PN}\,H_{\rm Newt})\,.
\end{eqnarray}
Using Eqs.~(\ref{HADM}) and (\ref{HADMothers}), we find that through 3.5PN order the SO couplings of the effective Hamiltonian are
\begin{equation}
\label{HeffSO}
H^{\rm eff}_{\rm SO} =\frac{1}{c^3}\,\frac{\boldsymbol{L}}{r^3}\cdot 
(g^{\rm eff}_{\sigma} \,\boldsymbol{\sigma}+g^{\rm eff}_{\sigma^\ast}\,\boldsymbol{\sigma^\ast})\,,
\end{equation}
where 
\begin{widetext}
\begin{subequations}
\label{gyroeffSO}
\begin{eqnarray}
g^{\rm eff}_\sigma&=& 2 + \frac{1}{c^2}\,
\left [ \frac{1}{8}(3\eta + 8a_0)\,\boldsymbol{\hat{p}}^2 - \frac{1}{2} (9 \eta + 6 a_0)\,(\boldsymbol{n} \cdot \boldsymbol{\hat{p}})^2 - (\eta + a_0)\,\frac{M}{r} \right ] \nonumber \\
&& +\frac{1}{c^4}\, \left[ \frac{1}{2}(-4a_0-2a_2-18\eta-a_0\eta-3\eta^2)\,\left(\frac{M}{r}\right)^2 
+ \frac{1}{8}(-4a_0+8a_1-5\eta-2a_0\eta)\,\boldsymbol{\hat{p}}^{4} +\frac{1}{8}(-4a_0 \right . \nonumber \\
&& \left. -8a_1 +8a_2-34\eta+6a_0\eta+11\eta^2)\,\frac{M}{r}\,\boldsymbol{\hat{p}}^{2} + \frac{3}{16}(8a_0-16a_1+16a_3+12\eta -20a_0\eta -13\eta^2)\,\boldsymbol{\hat{p}}^{2} \,(\boldsymbol{n} \cdot \boldsymbol{\hat{p}})^2 \right. \nonumber \\
&& \left. +\frac{1}{16}(32a_0-32a_1-64a_2-48a_3+140\eta+48a_0\,\eta-3\eta^2)\, \frac{M}{r}\,(\boldsymbol{n} \cdot \boldsymbol{\hat{p}})^2 + \frac{5}{16}(-16a_3+24a_0\eta \right .\nonumber \\
&& \left. + 27\eta^2)\,(\boldsymbol{n} \cdot \boldsymbol{\hat{p}})^4 \right ]\,, \\
g^{\rm eff}_{\sigma^\ast} &=&\frac{3}{2} +
\frac{1}{c^2}\, \left [ \frac{1}{8}(-5+ 4\eta + 8b_0)\,
\boldsymbol{\hat{p}}^2 - \frac{1}{4} (15\eta+12 b_0)\,(\boldsymbol{n} \cdot \boldsymbol{\hat{p}})^2 
- \frac{1}{4}(2+ 5 \eta + 4b_0)\,\frac{M}{r} \right ] \nonumber \\
&& +\frac{1}{c^4}\, \left[\frac{1}{8}(-4-16b_0-8b_2-55\eta-4b_0\eta-13\eta^2)\,\left(\frac{M}{r}\right)^2 +\frac{1}{16}(7-8b_0+16b_1-11\eta
-4b_0\eta-\eta^2)\,\boldsymbol{\hat{p}}^{4} 
\right. \nonumber \\
&& \left. +\frac{1}{16}(4 - 8b_0-16b_1+16b_2-59\eta+12b_0\eta+24\eta^2)\,\frac{M}{r}\,\boldsymbol{\hat{p}}^{2} + \frac{3}{16}(8b_0-16b_1+16b_3+19\eta-20b_0\eta \right. \nonumber \\
&& \left . -14 \eta^2)\,\boldsymbol{\hat{p}}^{2} \,(\boldsymbol{n} \cdot \boldsymbol{\hat{p}})^2 +\frac{1}{8}(10 +16b_0-16b_1-32b_2-24b_3+109\eta+24b_0\eta+6\eta^2)\, \frac{M}{r}\,(\boldsymbol{n} \cdot \boldsymbol{\hat{p}})^2 + \frac{5}{2}(-2b_3 \right. 
\nonumber \\
&& \left. +3b_0\,\eta +3\eta^2)\,(\boldsymbol{n} \cdot \boldsymbol{\hat{p}})^4 \right ]\,. 
\end{eqnarray}
\end{subequations}
\end{widetext}

\subsection{The PN-expanded  Hamiltonian of a spinning test-particle in a deformed Kerr spacetime}
\label{sec:PNexp}

The deformed-Kerr metric was obtained in Ref.~\cite{Barausse:2009xi}, and it reads
\begin{subequations}
\begin{eqnarray}
\label{def_metric_in}
g^{tt} &=& -\frac{\Lambda_t}{\Delta_t\,\Sigma}\,,\\
g^{rr} &=& \frac{\Delta_r}{\Sigma}\,,\\
g^{\theta\theta} &=& \frac{1}{\Sigma}\,,\\
g^{\phi\phi} &=& \frac{1}{\Lambda_t}
\left(-\frac{\tilde{\omega}_{\rm fd}^2}{\Delta_t\,\Sigma}+\frac{\Sigma}{\sin^2\theta}\right)\,,\label{eq:gff}\\
g^{t\phi}&=&-\frac{\tilde{\omega}_{\rm fd}}{\Delta_t\,\Sigma}\,.\label{def_metric_fin}
\end{eqnarray}
\end{subequations}
The potentials in these equations are given by
\begin{eqnarray}
\label{deltat}
\Delta_t &=& r^2\, \left [A(r) + \frac{a^2}{r^2} \right ]\,, \\
\label{deltar}
\Delta_r &=& \Delta_t\, D^{-1}(r)\,,\\
\Lambda_t &=& (r^2+a^2)^2 - a^2\,\Delta_t\,\sin^2\theta \,,\\
\Sigma &=& r^2+a^2\,\cos \theta^2\,,
\end{eqnarray}
and
\begin{equation}
\tilde{\omega}_{\rm fd} = 2a\,M\,r + a\,\eta\,\omega_{\rm fd}^0\,M^2+ a\,\eta\,\omega_{\rm fd}^1\,\frac{M^3}{r}\,,
\label{eq:omegafd}
\end{equation}
where $\omega_{\rm fd}^0$ and $\omega_{\rm fd}^1$ are two ``frame-dragging'' parameters (that we will fix later), and where
\begin{subequations}
\begin{align}
\label{A_PN}
& A(r) = 1 - \frac{2M}{r}  + \frac{2 \eta\,M^3}{r^3} + \left (\frac{94}{3} - \frac{41}{32} \pi^2\right)\, \frac{\eta\,M^4}{r^4}\,,
\\
& D^{-1}(r) = 1 + \frac{6 \eta\,M^2}{r^2} + 2 (26 - 3 \eta)\,\frac{\eta\,M^3}{r^3}\,.\label{D_PN}
\end{align}
\end{subequations}
The Hamiltonian of a spinning test-particle in the deformed-Kerr spacetime is
\begin{equation}
\label{Htotal}
H = H_{\rm NS} + H_{\rm S}\,,
\end{equation}
with 
\begin{equation}
\label{eq:Hnsdef}
{H}_{\rm NS} = \beta^i \, p_i + \alpha \sqrt{\mu^2 + \gamma^{ij}\,p_i\,p_j + {\cal Q}_4(p) }\,,
\end{equation} 
where the term ${\cal Q}_4(p)$ is quartic in the space momenta $p_i$ and 
was introduced in Ref.~\cite{2000PhRvD..62h4011D}. Moreover, we have 
\begin{eqnarray}
\label{alpha}
\alpha &=& \frac{1}{\sqrt{-g^{tt}}}\,,\\
\beta^i &=& \frac{g^{ti}}{g^{tt}}\,,\\
\gamma^{ij} &=& g^{ij}-\frac{g^{ti}\,g^{tj}}{g^{tt}}\,.
\label{gamma}
\end{eqnarray}
and 
\begin{align}
\label{KerrHS}
&H_{\rm S} = H_{\rm SO} + H_{\rm SS}\,,\\\nonumber&
\end{align}
where $H_{\rm SO}$ contains the odd terms in the spins (and therefore, in particular, the SO terms) 
and $H_{\rm SS}$ contains the  even terms in the spins (and therefore, in particular, the spin-spin terms
of the kind $S_{\rm Kerr} S^\ast$).

Since here we are interested in the SO couplings, we consider only $H_{\rm SO}$ ($H_{\rm SS}$ can be 
read from Eq.~(4.19) in Ref.~\cite{Barausse:2009xi}):
\begin{widetext}
\begin{eqnarray}
\label{HSOpartial}
  H_{\rm SO} &=& \frac{e^{2 \nu -\tilde{\mu} }\,\left(e^{\tilde{\mu} +\nu }-\tilde{B}\right)\, 
    (\boldsymbol{\hat{p}}\cdot \boldsymbol{\xi}\, r)\,  (\boldsymbol{S}\cdot \boldsymbol{\hat{S}}_{\rm Kerr})}{\tilde{B}^2\, \sqrt{Q}\,\xi^2} + 
   \frac{e^{\nu -2 \tilde{\mu} }}{\tilde{B}^2\, \left(\sqrt{Q}+1\right)\, \sqrt{Q}\, \xi^2}\Bigg\{
  (\boldsymbol{S}\cdot \boldsymbol{\xi})\, \tilde{J}
  \left[\mu_r\, (\boldsymbol{\hat{p}}\cdot \boldsymbol{v}\, r) \left(\sqrt{Q}+1\right) \right. \nonumber \\
&& \left. -\mu_{\cos\theta}\,(\boldsymbol{\hat{p}}\cdot \boldsymbol{n})\, 
    \xi^2 -\sqrt{Q}\, (\nu_r\, (\boldsymbol{\hat{p}}\cdot \boldsymbol{v}\, r)+(\mu_{\cos\theta}-\nu_{\cos\theta})\,
    (\boldsymbol{\hat{p}}\cdot \boldsymbol{n}) \,\xi^2)\right]\, \tilde{B}^2 +e^{\tilde{\mu} +\nu }\, 
  (\boldsymbol{\hat{p}}\cdot \boldsymbol{\xi}\, r)\, 
  \left(2 \sqrt{Q}+1\right)\,\times \nonumber \\
&& \Big[\tilde{J}\, \nu_r\, (\boldsymbol{S}\cdot \boldsymbol{v}) -\nu_{\cos\theta}\,
  (\boldsymbol{S}\cdot \boldsymbol{n})\, \xi^2\Big]\, \tilde{B} -\tilde{J}\, \tilde{B}_r\, e^{\tilde{\mu} +\nu }\, 
(\boldsymbol{\hat{p}}\cdot \boldsymbol{\xi}\, r)\, \left(\sqrt{Q}+1\right)\, (\boldsymbol{S}\cdot \boldsymbol{v})\Bigg\}\,,
\end{eqnarray}
\end{widetext}
where $\boldsymbol{\hat{S}}_{\rm Kerr}=\boldsymbol{{S}}_{\rm Kerr}/{{S}}_{\rm Kerr}$, $\boldsymbol{\xi} = \boldsymbol{\hat{S}}_{\rm Kerr}\times\boldsymbol{n}$,  
$\boldsymbol{v} = \boldsymbol{n} \times \boldsymbol{\xi}$, and where
\begin{equation}\label{Qpert}
Q=1+\frac{\Delta_r (\boldsymbol{\hat{p}}\cdot \boldsymbol{n})^2}{\Sigma}+
\frac{(\boldsymbol{\hat{p}}\cdot \boldsymbol{\xi}\, r)^2 \Sigma }{\Lambda_t\,\sin^2\theta}+
\frac{(\boldsymbol{\hat{p}}\cdot \boldsymbol{v}\, r)^2 }{\Sigma\,\sin^2\theta}\,,
\end{equation}
and
\begin{subequations}
\label{quantities}
\begin{align}
&\nu_r = \frac{r}{{\Sigma}}+\frac{(r^2+a^2)\left[(r^2+a^2)\,{\Delta^{\prime}_t}-4 r\, 
{\Delta_t}\right]}{2 {\Lambda_t}\, {\Delta_t}}\,,\\
&\nu_{\cos\theta}=\frac{a^2\, (r^2+a^2)\, \cos \theta\, ({r^2+a^2}-{\Delta_t})}{{\Lambda_t}\,
{\Sigma}}\,,\\
&\mu_r =\frac{r}{{\Sigma}}-\frac{1}{\sqrt{{\Delta_r}}}\,,\quad 
\mu_{\cos\theta}=\frac{a^2\, \cos\theta}{\Sigma}\,,\\
&\tilde{B} = \sqrt{\Delta_t}\,, \quad \tilde{B}_r = \frac{\sqrt{{\Delta_r}}\,{\Delta^\prime_t}-2 {\Delta_t}}
{2 \sqrt{{\Delta_r}\,{\Delta_t}}}\,,\\
&e^{2\tilde{\mu}} =\Sigma\,, \quad 
e^{2 \nu} = \frac{\Delta_t\,\Sigma}{\Lambda_t}\,, \quad \tilde{J} = \sqrt{\Delta_r}\,,
\end{align}
\end{subequations}
in which we use a prime to denote the derivative with respect to $r$. 
To obtain the SO couplings through 3.5PN order, we expand Eq.~(\ref{Htotal}). In particular, 
it is sufficient to consider the first term in the right-hand-side of 
Eq.~(\ref{eq:Hnsdef}), and set $a=0$ (deformed-Schwarzschild limit) in 
Eqs.~(\ref{HSOpartial}), (\ref{Qpert}) and (\ref{quantities}). Doing so, 
for the PN-expanded deformed-Kerr Hamiltonian we obtain
\begin{subequations}
\label{HNSPNd}
\begin{align}
&H^{\rm NS}_{\rm SO\,1.5PN} = \frac{2}{r^3\,c^3}\, \boldsymbol{L}\cdot \boldsymbol{S}_{\rm Kerr} \,,\\
&H^{\rm NS}_{\rm SO\,2.5PN} = \frac{1}{r^3\,c^5}\,\eta\,\omega_{\rm fd}^0\,\frac{M}{r}\,\boldsymbol{L}\cdot \boldsymbol{S}_{\rm Kerr}\,,\\
&H^{\rm NS}_{\rm SO\,3.5PN} = \frac{1}{r^3\,c^7}\,\eta\,\omega_{\rm fd}^1\,\left (\frac{M}{r}\right)^2\,
\boldsymbol{L}\cdot \boldsymbol{S}_{\rm Kerr} \,,
\end{align}
\end{subequations}
and
\begin{subequations}
\label{HSPNd}
\begin{align}
  &{H}^{\rm S}_{\rm SO\,1.5PN} = \frac{3}{2r^3\,c^3} \, \boldsymbol{L}\cdot {\boldsymbol{S}}^\ast\,,
  \label{eq:H15PN}\\
&{H}^{\rm S}_{\rm SO\,2.5PN} = \frac{1}{r^3\,c^5}\, \boldsymbol{L}\cdot\boldsymbol{S}^\ast\,\left[-\frac{1}{2}(1 + 6\eta)\,
\frac{M}{r} - \frac{5}{8} \boldsymbol{\hat{p}}^2
  \right]\,, \\
&{H}^{\rm S}_{\rm SO\,3.5PN} = \frac{1}{r^3\,c^7}\,\boldsymbol{L}\cdot {\boldsymbol{S}}^\ast\,
\left[  \frac{1}{2}(-1 - 42 \eta + 6 \eta^2)\, \left(\frac{M}{r}\right)^2 \right. \nonumber \\
& \left. + \frac{7}{16}\,\boldsymbol{\hat{p}}^4 + \frac{1+6\eta}{4} \left(\frac{M}{r}\right)\,\boldsymbol{\hat{p}}^2 
+ \frac54 \left(\frac{M}{r}\right)\,(\boldsymbol{n} \cdot \boldsymbol{\hat{p}} )^2\right]\,. 
\end{align}
\end{subequations}

\subsection{The EOB Hamiltonian:  spin-mapping dependent on dynamical  variables}
\label{sec:SEOBgeo}

We now determine the mapping between the spins $\boldsymbol{\sigma}$ and $\boldsymbol{\sigma}^\ast$ 
of the effective ADM Hamiltonian and the spins $\boldsymbol{S}_{\rm Kerr}$ and $\boldsymbol{S}^\ast$ 
of the EOB Hamiltonian by imposing that the deformed-Kerr Hamiltonian given by  
Eqs.~(\ref{HNSPNd}) and (\ref{HSPNd}) coincides with the effective Hamiltonian given by 
Eqs.~(\ref{HeffSO}) and (\ref{gyroeffSO}). As found in Ref.~\cite{Barausse:2009xi}, we have to assume that
the mapping depends on the orbital dynamical variables $\mathbf{p}^2$, $\mathbf{n} \cdot \mathbf{p}$  
and $r$. The general mapping of the spins has the form
\begin{subequations}
\begin{eqnarray}
\boldsymbol{{S}}^\ast &=& \boldsymbol{\sigma}^\ast+\frac{1}{c^2}\,\boldsymbol{\Delta}^{(1)}_{{\sigma}^\ast}
+\frac{1}{c^4}\,\boldsymbol{\Delta}^{(2)}_{{\sigma}^\ast}\,,\\
\boldsymbol{{S}}_{\rm Kerr} &=& \boldsymbol{\sigma}+\frac{1}{c^2}\,\boldsymbol{\Delta}^{(1)}_{{\sigma}} 
+\frac{1}{c^4}\,\boldsymbol{\Delta}^{(2)}_{{\sigma}}\,.
\end{eqnarray}
\end{subequations}
At 2.5PN order, if we assume $\omega^0_{\rm fd}=0$ [see Eq.~\eqref{eq:omegafd}] and $\boldsymbol{\Delta}^{(1)}_{{\sigma}}=0$, 
we have~\cite{Barausse:2009xi}  
\begin{eqnarray}\label{deltaSigma1}
\boldsymbol{\Delta}^{(1)}_{\sigma^\ast}&=&\boldsymbol{\sigma}^\ast\,  
\left [ \frac{1}{6}(-4b_0+7\eta)\,\frac{M}{r} + \frac{1}{3}(2b_0+\eta)\, (Q-1) \right. \nonumber \\
&& \left. -\frac{1}{2}(4b_0+5\eta)\, \frac{\Delta_r}{\Sigma}\,
(\boldsymbol{n}\cdot \boldsymbol{\hat{p}})^2 \right ] \nonumber \\
&& + \boldsymbol{\sigma}\,\left [ -\frac{2}{3}(a_0+\eta)\,\frac{M}{r} + \frac{1}{12}(8a_0+3\eta)\,(Q-1)
\right. \nonumber \\
&& \left. -(2a_0+3\eta)\,\frac{\Delta_r}{\Sigma}\,(\boldsymbol{n}\cdot \boldsymbol{\hat{p}})^2 \right ]\,,
\end{eqnarray}
and at  3.5PN order, assuming $\omega_{\rm fd}^1=0$ [see Eq.~\eqref{eq:omegafd}] 
and $\boldsymbol{\Delta}^{(2)}_{{\sigma}}=0$, we obtain
\begin{widetext}
\begin{eqnarray}\label{deltaSigma2}
\boldsymbol{\Delta}^{(2)}_{{\sigma}^\ast}&=& \boldsymbol{\sigma}^\ast\,
\left [\frac{1}{36}\,(-56b_0-24b_2+353 \eta-60b_0\eta-27\eta^2)\,\left(\frac{M}{r}\right)^2 +\frac{5}{3}(-2b_3+3b_0\eta+3\eta^2)\,
\frac{\Delta_r^2}{\Sigma^2}\,(\boldsymbol{n} \cdot \boldsymbol{\hat{p}})^4 +\frac{1}{72}\,(-4b_0+48b_1\right. \nonumber \\
&& \left. -23\eta -12b_0\eta -3\eta^2)\,(Q-1)^2 +\frac{1}{36}\,(-14b_0-24b_1+24b_2-103\eta+66b_0\eta+60\eta^2)\,\frac{M}{r}\,(Q-1) 
+\frac{1}{12}\,(2b_0-24b_1 \right. \nonumber \\
&& \left. +24b_3 +16\eta -30b_0\eta-21\eta^2)\,\frac{\Delta_r}{\Sigma}\,(\boldsymbol{n} \cdot \boldsymbol{\hat{p}})^2\,(Q-1)
+\frac{1}{12}\,(-24b_0-16b_1-32b_2-24b_3+ 47\eta-24b_0\eta-54\eta^2)\,\times \right. \nonumber \\
&& \left. \frac{M}{r}\,\frac{\Delta_r}{\Sigma}\,(\boldsymbol{n} \cdot \boldsymbol{\hat{p}})^2 \right ] \nonumber \\
&& + \boldsymbol{\sigma} \, \left [ \frac{1}{9}\,(-14a_0 -6a_2-56\eta-15a_0\eta-21\eta^2)\,\left(\frac{M}{r}\right)^2 
+\frac{5}{24}(-16a_3+24a_0\eta+27\eta^2)\,\frac{\Delta_r^2}{\Sigma^2}\,(\boldsymbol{n} \cdot \boldsymbol{\hat{p}})^4 +
\frac{1}{144}(-8a_0   \right. \nonumber \\
&& \left. +96a_1  -45 \eta -24a_0\eta)\,(Q-1)^2  
+\frac{1}{36}\,(-14a_0-24a_1+24 a_2-109\eta +66a_0\eta+51\eta^2)\,\frac{M}{r}\,(Q-1) + \frac{1}{24}\,(4 a_0
\right. \nonumber \\
&& \left. -48a_1 + 48a_3 +6 \eta - 60a_0\eta - 39\eta^2)\,\frac{\Delta_r}{\Sigma}\,(\boldsymbol{n} \cdot \boldsymbol{\hat{p}})^2\,(Q-1) 
+\frac{1}{24}\,(-48a_0-32a_1-64a_2-48a_3-16\eta -48a_0\eta \right. \nonumber \\
&& \left. -147\eta^2)\,\frac{M}{r}\,\frac{\Delta_r}{\Sigma}\,(\boldsymbol{n} \cdot \boldsymbol{\hat{p}})^2 \right ]\,. 
\end{eqnarray}
\end{widetext}
Note that as in Ref.~\cite{Barausse:2009xi}, we have replaced, in the expressions for
$\boldsymbol{\Delta}^{(1)}_{{\sigma}^\ast}$ and $\boldsymbol{\Delta}^{(2)}_{{\sigma}^\ast}$, the term $\boldsymbol{\hat{p}}^2$ with 
$\gamma^{ij}\hat{p}_i\hat{p}_j=Q-1$ and  the term $(\boldsymbol{n} \cdot \boldsymbol{\hat{p}})^2$ with  
$\Delta_r(\boldsymbol{n} \cdot \boldsymbol{\hat{p}})^2/\Sigma=g^{rr}\hat{p}_r^2$. 

Having determined the spin mappings, we can write down the real improved (or EOB) Hamiltonian 
for spinning black holes, which turns out to be
\begin{equation}
\label{hreal}
H_\mathrm{real}^{\rm improved} = M\,\sqrt{1+2\eta\,\left(\frac{H_{\rm eff}}{\mu}-1\right)}\,,
\end{equation}
where
\begin{eqnarray}
\label{HeffEOB-geo}
H_{\rm eff} &=& {H}_{\rm S}+ \beta^i\,p_i+ \alpha\, \sqrt{\mu^2 + \gamma^{ij}\,p_i\,p_j + {\cal Q}_4(p)} \nonumber \\
&& -\frac{\mu}{2M\, r^3}\,(\delta^{ij} - 3 n^i\,n^j)\, S^\ast_i\, S^\ast_j\,.
\nonumber \\
\end{eqnarray}

\subsection{The EOB Hamiltonian:  spin-mapping independent of dynamical variables}
\label{sec:SEOBngeo}

In the previous section, we had to assume a dependence
on the orbital dynamical variables $\mathbf{p}^2$, $\mathbf{n} \cdot \mathbf{p}$  and $r$
in the mapping between 
the spins $\boldsymbol{\sigma}$ and $\boldsymbol{\sigma}^\ast$
of the effective ADM Hamiltonian and the spins $\boldsymbol{S}_{\rm Kerr}$ and $\boldsymbol{S}^\ast$ 
of the deformed-Kerr Hamiltonian. 
To avoid this dependence on the dynamical 
variables and obtain the much simpler mapping 
\begin{subequations}
\begin{eqnarray}
\boldsymbol{{S}}^\ast &=& \boldsymbol{\sigma}^\ast\,,\\
\boldsymbol{{S}}_{\rm Kerr} &=& \boldsymbol{\sigma}\,,
\end{eqnarray}
\end{subequations}
we need to modify the Hamilton-Jacobi equation by adding terms depending on the momenta 
and spins. Since in this paper we are dealing only with SO effects, we will neglect
modifications that involve spin-spin terms. We start by observing that in the presence of spins 
the linear momentum $P_\mu$, which is related to the canonical momentum by $P_\mu = p_\mu + E^{\rho\sigma}_\mu S^\ast_{\rho\sigma}$ 
[see Eq.~(3.28) of Ref.~\cite{Barausse:2009aa}], satisfies the Hamilton-Jacobi equation~\cite{Barausse:2009aa}
\begin{equation}\label{unpert_HJ}
\mu^2 + P_\mu\,P^\mu =\mu^2 + p_\mu\,p^\mu+2 E^{\rho \sigma}_\mu  p^\mu S^\ast_{\rho\sigma}+{\cal O} (S^\ast)^2=0\,.
\end{equation}
Here, $S^\ast_{\mu \nu}$ is the spin tensor of the test-particle 
[see Ref.~\cite{Barausse:2009aa} and also  Eqs.~(2.4)--(2.7) in Ref.~\cite{Faye-Blanchet-Buonanno:2006}]. 
Equation~\eqref{unpert_HJ} leads to the correct Hamiltonian
for a spinning particle in curved spacetime, at linear order in the particle's spin~\cite{Barausse:2009aa}.
To modify the Hamilton-Jacobi equation, a suitable ansatz is
\begin{align}
\label{eqHJgen}
\mu^2 &  + g_{\rm eff}^{\mu \nu}(\mathbf{r},\mathbf{S}_{\rm Kerr})\,p_\mu\,p_\nu 
+2 E^{\rho \sigma}_\mu p^\mu S^\ast_{\rho\sigma} \nonumber \\
&
+  [B^{\mu \nu \lambda}_{\rho \sigma}(\mathbf{r})\,p_\mu\,p_\nu\,p_\lambda
\nonumber \\
& + B^{\mu \nu \lambda \tau \alpha}_{\rho \sigma}(\mathbf{r})\,p_\mu\,p_\nu\,p_\lambda\,p_\tau\,p_\alpha]\,S_\ast^{\rho \sigma} 
\nonumber \\
& +   A^{\mu \nu \lambda \tau}(\mathbf{r},\mathbf{S}_{\rm Kerr})\,p_\mu\,p_\nu\,p_\lambda\,p_\tau \nonumber \\
& +   A^{\mu \nu \lambda \tau\rho\sigma}(\mathbf{r},\mathbf{S}_{\rm Kerr})\,p_\mu\,p_\nu\,p_\lambda\,p_\tau p_\rho\,p_\sigma \nonumber \\ 
&+ \cdots = 0\,.
\end{align}
If we make use at lowest order of the condition $\mu^2 + 
g_{\rm eff}^{\mu \nu}\,p_\mu\,p_\nu  \simeq 0$, 
we can replace $p_t$ with the spatial components of the momentum, and obtain 
the following generalized form of the effective Hamiltonian
\begin{align}
\label{HeffEOB-ngeo}
&H_{\rm eff} =\nonumber \\&\beta^i\,p_i+ \alpha\, \sqrt{\mu^2 + \gamma^{ij}\,p_i\,p_j + {\cal Q}_4(p) + 
{\cal Q}_S(r,p,S^\ast,S_{\rm Kerr})} \nonumber \\
& -\frac{\mu}{2M\, r^3}\,
(\delta^{ij} - 3 n^i\,n^j)\, S^\ast_i\, S^\ast_j+ {H}_{\rm S}\,,\\\nonumber
\end{align}
where ${\cal Q}_4(p)$ is a quartic term in the momenta~\cite{2000PhRvD..62h4011D},
which is due to the presence of the quartic term $A^{\mu \nu \lambda \tau}\,p_\mu\,p_\nu\,p_\lambda\,p_\tau$ in Eq.~\eqref{eqHJgen},
and ${\cal Q}_S(r,p,S^\ast,S_{\rm Kerr})$ is 
a term linear in $\boldsymbol{S}_\ast$ and $\boldsymbol{S}_{\rm Kerr}$
\begin{equation}
{\cal Q}_S(r,p,S^\ast,S_{\rm Kerr})={\cal Q}_i^{S_{\rm Kerr}}(r,p)S^i_{\rm Kerr}+{\cal Q}_i^{S^\ast}(r,p)S^i_{\ast}\,.
\end{equation}
In particular, the term ${\cal Q}_i^{S^\ast}(r,p)S^i_{\ast}$ comes
from the terms $B^{\mu \nu \lambda}_{\rho \sigma}(\mathbf{r})\,p_\mu\,p_\nu\,p_\lambda S_\ast^{\rho \sigma}$ 
and $B^{\mu \nu \lambda \tau \alpha}_{\rho \sigma}\,p_\mu\,p_\nu\,p_\lambda\,p_\tau\,p_\alpha S_\ast^{\rho \sigma}$
in Eq.~\eqref{eqHJgen}, while the term ${\cal Q}_i^{S_{\rm Kerr}}(r,p)S^i_{\rm Kerr}$ comes from 
$ A^{t \nu \lambda \tau}\,p_t\,p_\nu\,p_\lambda\,p_\tau$ and
$ A^{t \nu \lambda \tau\rho\sigma}\,p_t\,p_\nu\,p_\lambda\,p_\tau p_\rho\,p_\sigma$ (through the dependence
of the tensors $A^{\mu \nu \lambda \tau}$ and $A^{\mu \nu \lambda \tau\rho\sigma}$ on $\boldsymbol{S}_{\rm Kerr}$). 
Finally, the term $H_{\rm S}$ in Eq.~\eqref{HeffEOB-ngeo} comes from the presence of $2 E^{\rho \sigma}_\mu p^\mu S^\ast_{\rho\sigma}$ 
in Eq.~\eqref{eqHJgen}. As already stressed, this happens because Eq.~\eqref{unpert_HJ} leads to the correct Hamiltonian
for a spinning particle in curved spacetime, and in particular to $H_{\rm S}$, which is the spin-dependent part of
that Hamiltonian~\cite{Barausse:2009aa}.

Through 3.5PN order the quantities ${\cal Q}_i^{S_{\rm Kerr}}(r,p)\,S^i_{\rm Kerr}$ and ${\cal Q}_i^{S^\ast}(r,p)\,S^i_{\ast}$ must have the form
\begin{align}
{\cal Q}^s_i(r,p)s^i &= \frac{\mu}{r^2\,c^3}\,\epsilon_{i j k} n^j\, p^k s^i\times\nonumber\\& \left \{ 
\frac{1}{c^2} \left (c_1\frac{M}{r} + c_2\,\p^2 + c_3\,\np^2 \right ) \right. \nonumber \\
& \left. + \frac{1}{c^4} \left [c_4\,\p^4 + c_5\left (\frac{M}{r}\right)^2  + c_6\,\np^4 +\right. \right.\nonumber \\
& \left. \left.
c_7\,\p^2\,\frac{M}{r} + c_8\np^2\,\frac{M}{r} + c_9\,\np^2\,\p^2 \right ] \right \}\,,  
\end{align}
where $s$ stands for either $S_{\rm Kerr}$ or $S^\ast$, while the coefficients $c_n$, $n = 1, \dots 9$ are determined 
by the mapping of the effective to the real description. 
A straightforward computation leads to
\begin{eqnarray}
{\cal Q}_S= {\cal Q}_{S\, \rm 2.5PN}+ {\cal Q}_{S\, \rm 3.5PN}\,,
\end{eqnarray}
where
\begin{widetext}
\begin{eqnarray}
\label{SNG25}
 {\cal Q}_{S\,\rm 2.5PN}(\mathbf{r},\mathbf{p},\mathbf{S}^\ast,\mathbf{S}_{\rm Kerr}) &=& \frac{\mu}{r^3\,c^5}\, 
\left \{(\boldsymbol{{S}}_{\rm Kerr} \cdot \mathbf{L})\,\left [-2 (a_0+ \eta)\,\frac{M}{r} + \frac{1}{4}(8a_0+3\eta)\, (Q-1) 
-3 (2a_0+3\eta)\,\frac{\Delta_r}{\Sigma}\,
(\boldsymbol{n}\cdot \boldsymbol{\hat{p}})^2 \right ]\right.  \nonumber \\
&& \left . + (\boldsymbol{{S}}^\ast\cdot \mathbf{L})\,\left [\frac{1}{2} (-4b_0+ 7\eta)\,\frac{M}{r} 
+ (2b_0+\eta)\, (Q-1) -\frac{3}{2} (4b_0+5\eta)\, \frac{\Delta_r}{\Sigma}\,
(\boldsymbol{n}\cdot \boldsymbol{\hat{p}})^2 \right ]\right \}\,,
 \end{eqnarray}
\begin{eqnarray}
\label{SNG35}
 {\cal Q}_{S\,\rm 3.5PN}(\mathbf{r},\mathbf{p},\mathbf{S}^\ast,\mathbf{S}_{\rm Kerr}) &=&
 \frac{ \mu}{r^3\,c^7} \left\{ 
(\boldsymbol{{S}}_{\rm Kerr}\cdot \boldsymbol{L})\,
\left [(-6a_0 -2a_2 -20\eta - a_0\eta-3\eta^2)\,\left(\frac{M}{r}\right)^2 
+\frac{5}{8}(-16 a_3+24a_0\eta +27\eta^2)\,\times
\right. \right. \nonumber \\
&& \left. \left. \frac{\Delta_r^2}{\Sigma^2}\,(\boldsymbol{n} \cdot \boldsymbol{\hat{p}})^4 
+ \frac{1}{8}(16 a_1  -7 \eta -4a_0 \eta)\,(Q-1)^2  
+\frac{1}{4} (-8a_1+8 a_2-35\eta +6a_0 \eta+ 11\eta^2)\,\frac{M}{r}\,\times \right. \right. \nonumber \\
&& \left. \left. (Q-1)+ \frac{3}{8}\,(-16a_1+16 a_3 -20a_0\eta -13\eta^2)\,
\frac{\Delta_r}{\Sigma}\,(\boldsymbol{n} \cdot \boldsymbol{\hat{p}})^2\,(Q-1) 
+\frac{1}{8}(-80a_0-32a_1 \right. \right. \nonumber \\
&& \left. \left. -64a_2-48a_3-64\eta +48a_0\eta-3\eta^2)\, \frac{M}{r}\,\frac{\Delta_r}{\Sigma}\,(\boldsymbol{n} \cdot \boldsymbol{\hat{p}})^2 \right ] \right. \nonumber \\
&& + \left. (\boldsymbol{{S}}^\ast \cdot \boldsymbol{L})\,\left [\frac{1}{4}(-24b_0-8b_2+127\eta -4b_0\eta-37\eta^2)\,\left(\frac{M}{r}\right)^2 
+ 5(-2 b_3+3b_0\eta+3\eta^2)\,\times   \right. \right. \nonumber \\
&& \left. \left. 
\frac{\Delta_r^2}{\Sigma^2}\,(\boldsymbol{n} \cdot \boldsymbol{\hat{p}})^4 +\frac{1}{8}(16 b_1-7 \eta -4b_0\eta-\eta^2)\,(Q-1)^2  
+ \frac{1}{8}(-16b_1+16 b_2-61\eta  +12b_0\eta   \right. \right. \nonumber \\
&& \left. \left. + 24\eta^2)\,\frac{M}{r}\,(Q-1) + \frac{3}{8}(-16b_1+16 b_3 +9\eta-20b_0\eta-14\eta^2)\,
\frac{\Delta_r}{\Sigma}\,(\boldsymbol{n} \cdot \boldsymbol{\hat{p}})^2\,(Q-1) 
\right. \right. \nonumber \\
&& \left. \left. +\frac{1}{4}(-40b_0 -16b_1 -32b_2 -24b_3 +27\eta+24b_0\eta+6\eta^2)\, \frac{M}{r}\,\frac{\Delta_r}{\Sigma}\,(\boldsymbol{n} \cdot \boldsymbol{\hat{p}})^2 \right ] 
\right\}\,.
\end{eqnarray}
\end{widetext}
Finally, the EOB Hamiltonian is obtained by inserting Eq.~(\ref{HeffEOB-ngeo}) into Eq.~(\ref{hreal}).

\section{The effective-one-body dynamics for equatorial orbits}
\label{sec:EOBdynamics}

We stress that the EOB models introduced in the previous sections have
the correct test-particle limit, for both non-spinning and spinning
black holes (for \textit{generic} orbits and \textit{arbitrary} spin
orientations), and that the test-particle limit is recovered
non-perturbatively, (i.e., at \textit{all} PN orders). This is because
in order to build our models, in Sec.~\ref{sec:PNexp} we started from
the Hamiltonian derived in Ref.~\cite{Barausse:2009aa}, which
correctly reproduces the Mathisson-Papapetrou-Pirani equation
describing the motion of a classical spinning particle in a generic
curved spacetime~\cite{Math,Papa51, Papa51spin, CPapa51spin,Pirani}.
The EOB models that we present in this paper share this feature with
our earlier model~\cite{Barausse:2009xi}, which was valid
through 3PN order in the non-spinning sector and through 2.5PN order in
the spinning sector, but not with other EOB models for spinning
black-hole binaries, which recover the test-particle limit only
approximately~\cite{DJSspin}.

\begin{figure}[!t]
\includegraphics[width=7cm, clip=true]{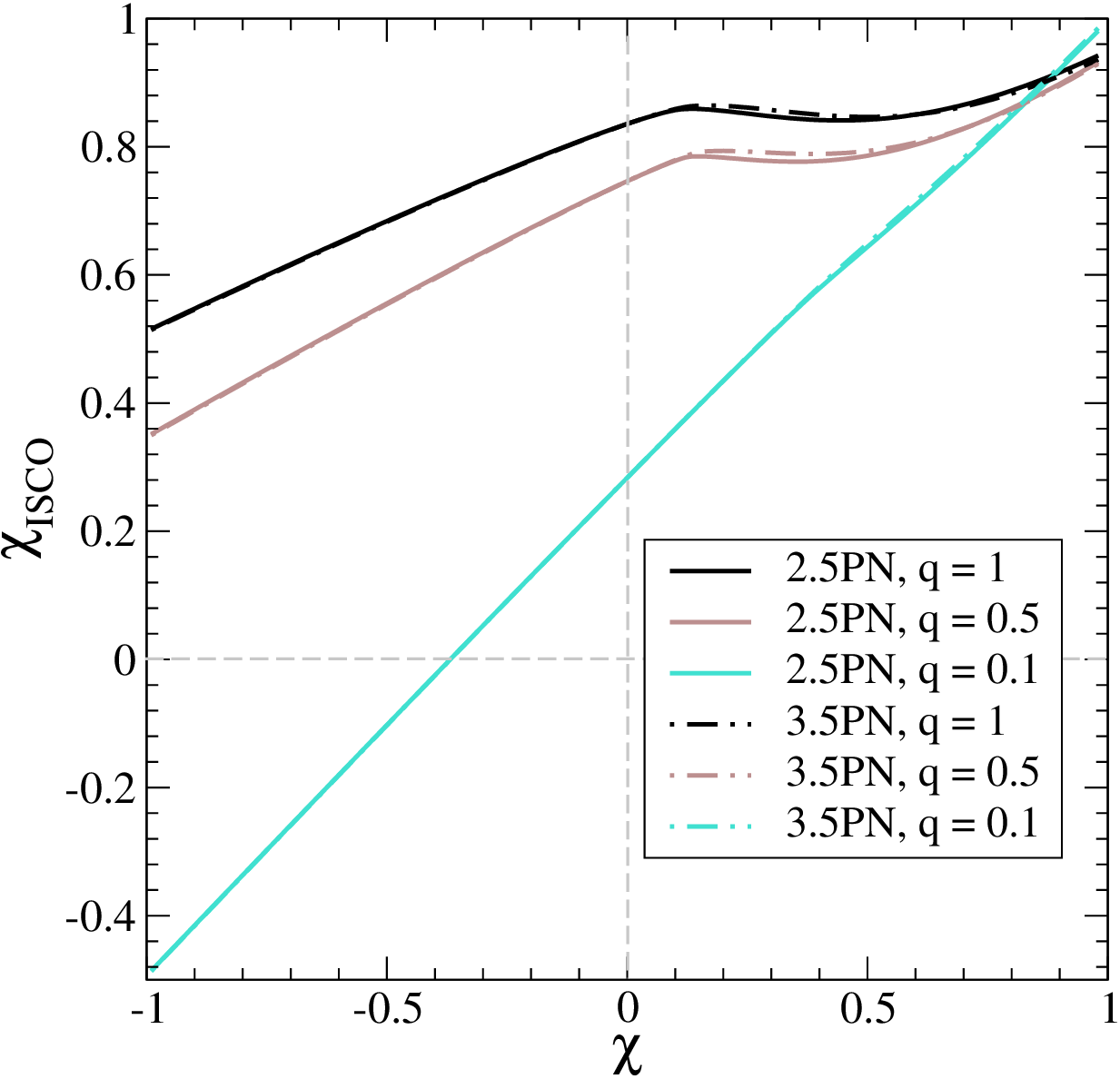}  
\caption{\label{fig:finalspin} The spin parameter of the binary
at the ISCO given by Eq.~\eqref{chiISCO} for the 2.5PN and 3.5PN EOB models with
dynamical mapping of the spins, for binaries having
spins parallel to $\boldsymbol{L}$, mass ratio $q = m_2/m_1$ and 
spin-parameter projections onto the direction of $\boldsymbol{L}$ given by
$\chi_1 = \chi_2 = \chi$.}
\end{figure}

Other attractive features of our models are evident when considering 
configurations with spins parallel to the orbital angular momentum, which correspond, in 
the effective EOB dynamics, to a particle  moving on equatorial orbits. For aligned spins and equatorial orbits, in fact,
both the models with dynamical and non-dynamical spin mapping 
predict the existence of an innermost stable circular orbits (ISCO), for all values of
the system's parameters. This feature is again shared by our earlier model~\cite{Barausse:2009xi}, 
but not by other EOB models for spinning black-hole binaries~\cite{DJSspin}, which do not present ISCOs for large values of the spins. While the non-existence of an ISCO is not necessarily a sign that a model is 
flawed, its presence helps reproduce the results of numerical-relativity simulations for binaries with aligned spins~\cite{Taracchini:2011}. 

To calculate the radius and the orbital angular momentum at the ISCO for our 
EOB models, we solve numerically the following system of equations~\cite{Buonanno00}
\begin{eqnarray}
&& \frac{\partial H^{\rm improved}_\mathrm{real}(r,p_r=0,L_z)}{\partial r}=0\,, \\
&& \frac{\partial^2 H^{\rm improved}_\mathrm{real}(r,p_r=0,L_z)}{\partial r^2}=0\,,
\end{eqnarray}
with respect to $r$ and $L_z=p_\phi$. The solutions can then be used to evaluate the ISCO
frequency via
\begin{equation}
\Omega_{_{\rm ISCO}}=\frac{\partial H^{\rm improved}_\mathrm{real}(r_{_{\rm ISCO}},p_r=0,L^{^{\rm ISCO}}_z)}{\partial L_z}\,,
\end{equation}
which follows immediately from the Hamilton equations. 

\begin{figure}
\includegraphics[width=7cm, clip=true]{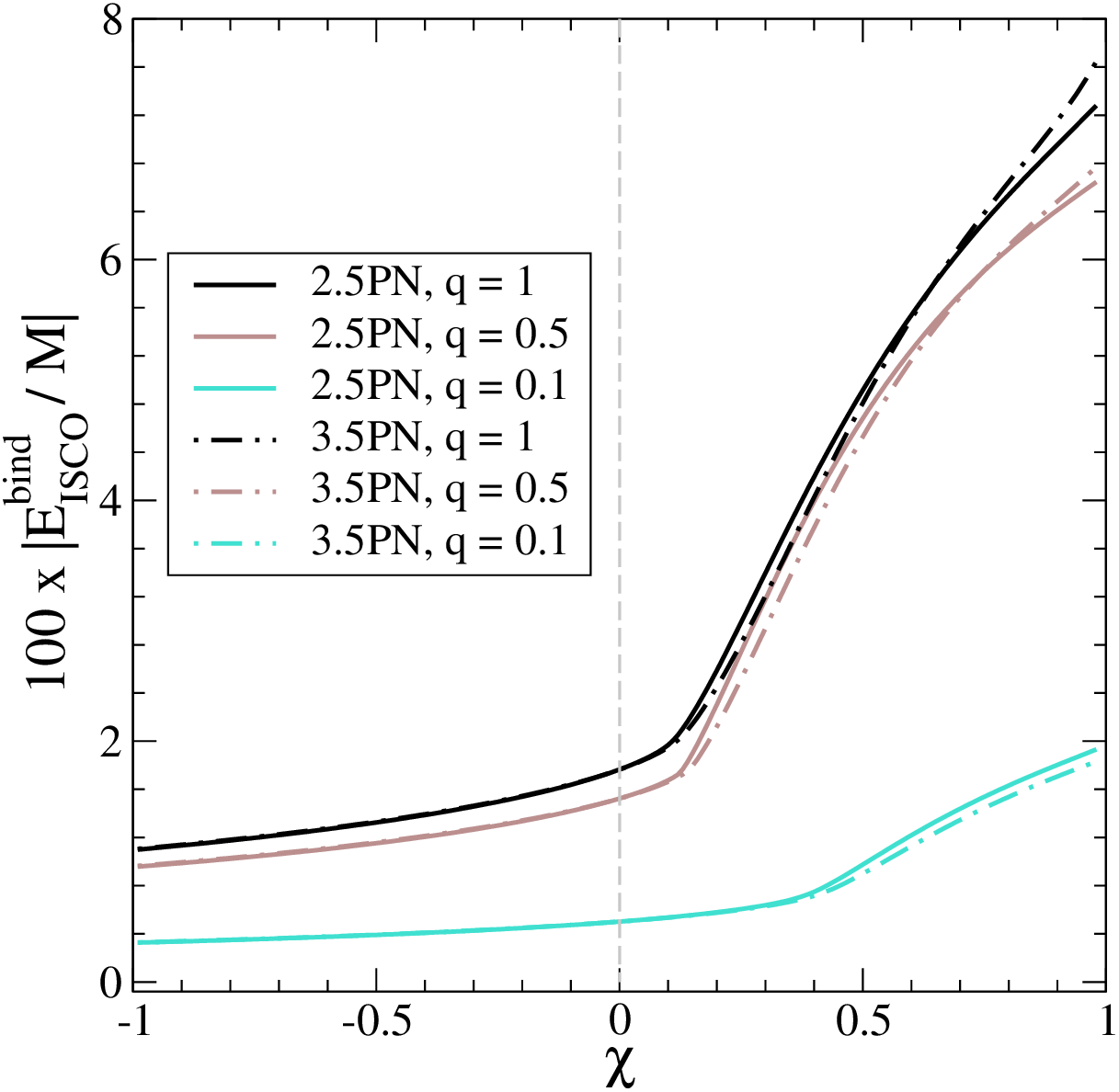}  
\caption{\label{fig:finalmass} The same as in Fig.~\ref{fig:finalspin} but for the binding energy of the binary at the ISCO.} 
\end{figure}

The values of $r_{_{\rm ISCO}}$ and $L^{^{\rm ISCO}}_z$ can also be used to calculate the binding energy at the ISCO via
$E_{\rm bind}=H^{\rm improved}_\mathrm{real}-M\,$. This quantity is interesting because it corresponds
to the mass lost in gravitational waves during the binary's inspiral, and is therefore a lower limit 
to the total mass loss, to which it reduces for $\eta\to0$ (when the fluxes during the merger and the 
ringdown become negligible~\cite{Barausse:2009xi}). Similarly, one can estimate the spin
of the binary at the ISCO via
\begin{equation}\label{chiISCO}
\chi_{_{\rm ISCO}}=\frac{S^z_1+S^z_2+L^z_{_{\rm ISCO}}}{(M+E^{\rm bind}_{_{\rm ISCO}})^2}\,.
\end{equation} 
This expression clearly neglects the mass and angular momentum lost during the merger and ringdown phases,
but it is useful as qualitative diagnostics of our model, and it reduces to the spin of the final black-hole remnant
when $\eta\to0$ (again, because in this limit the fluxes during the merger and the ringdown become negligible~\cite{Barausse:2009xi}).

We re-write the metric potentials $\Delta_t$ and $\Delta_r$ given in 
Eqs.~\eqref{deltat}, ~\eqref{deltar},  using the 
``log-model'' of Ref.~\cite{Barausse:2009xi} [see Eqs.~(5.71) 
and~(5.73)--(5.83) of that paper], and assume
\begin{equation}\label{Keta}
K(\eta)=1.447 - 0.1574\,\eta - 9.082\,\eta^2\,.
\end{equation} 
The value of $K(\eta)$ for $\eta=0$
ensures~\cite{Barausse:2009xi,favata} that the ISCO frequency for
extreme mass-ratio non-spinning binaries predicted by our EOB models
agrees with the exact result of Ref.~\cite{BarackSago09}, which
calculated the shift of the ISCO frequency due to the conservative
part of the self-force. The linear and quadratic terms in $\eta$ 
in Eq.~\eqref{Keta} are such that our EOB models accurately
reproduce numerical relativity simulations for non-spinning binaries
with mass ratios ranging from $q=1/6$ to $q=1$~\cite{Taracchini:2011}.

\begin{figure}[!t]
\includegraphics[width=7cm, clip=true]{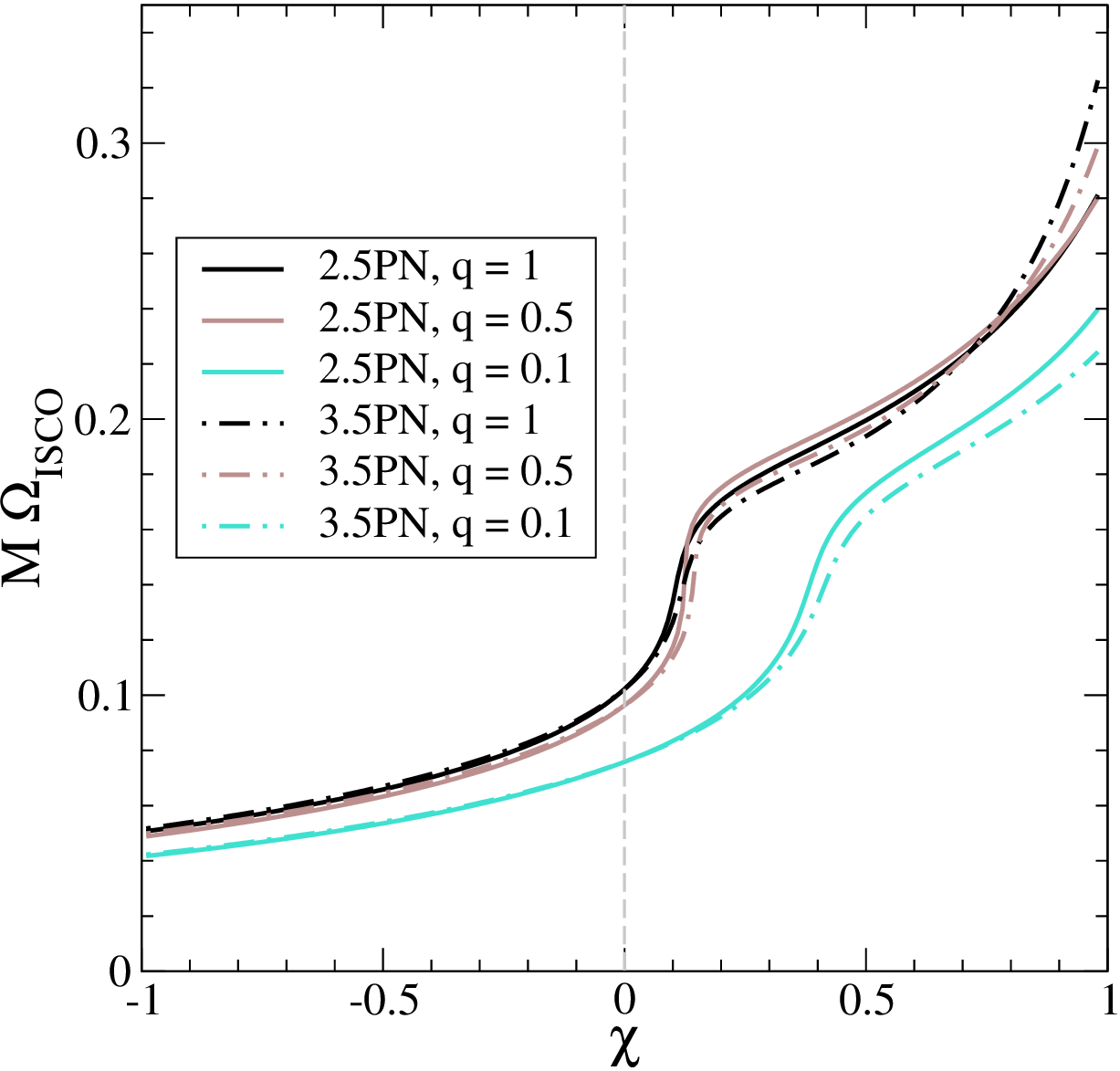}
\caption{\label{fig:omega-isco} The same as in Fig.~\ref{fig:finalspin} but for the ISCO frequency.}
\end{figure}

We fix the gauge parameters to the following values:
\begin{eqnarray}\label{gaugeparameters_start}
&& a_0 = -\frac{3}{2}\eta\,, \quad b_0 = - \frac{5}{4}\eta\,,\\
&& a_1 = \frac{1}{2}\eta^2\,, \quad b_1 = \frac{1}{16}\eta\,(9+5\eta)\,, \nonumber \\
&& a_2 = \frac{1}{8}\eta\,(7-8\eta)\,, \quad b_2 = \frac{1}{8}\eta\,(17-5 \eta)\,, \nonumber \\
&& a_3= -\frac{9}{16}\eta^2\,, \quad b_3 = -\frac{3}{8}\eta^2\,,\label{gaugeparameters_end}
\end{eqnarray}
which we determine by requiring that all the terms involving ${\Delta_r}\,\hat{\boldsymbol{p}}\cdot \boldsymbol{n}/\Sigma$
cancel out in $\boldsymbol{\Delta}^{(1)}_{{\sigma}^\ast}$ and
 $\boldsymbol{\Delta}^{(2)}_{{\sigma}^\ast}$ [Eqs.~\eqref{deltaSigma1}, ~\eqref{deltaSigma2}], 
or equivalently in ${\cal Q}_{S\,\rm 2.5PN}$ and ${\cal Q}_{S\,\rm 3.5PN}$ [Eqs.~\eqref{SNG25}, ~\eqref{SNG35}].
Different choices of the gauge parameters produce qualitatively similar results for the ISCO quantities that we described above.

Focusing on systems with spins aligned with the orbital angular
momentum $\boldsymbol{L}$, and denoting with
$S_{1,2}=\chi_{1,2}\,m^2_{1,2}$ the projections of the spins along the
direction of $\boldsymbol{L}$, we consider binaries with
$\chi_1=\chi_2=\chi$ and mass ratios $q=m_2/m_1=0.1$, $0.5$ and
$1$. In particular, in Figs.~\ref{fig:finalspin}--\ref{fig:omega-isco}
we show how the ISCO quantities described above change as a
consequence of including the 3.5PN SO terms in our EOB model with
dynamical spin mapping. More specifically, we calculate $\Omega_{_{\rm
    ISCO}} M$, $E^{\rm ISCO}_{\rm bind}/M$ and $\chi_{_{\rm ISCO}}$
using the Hamiltonian \eqref{HeffEOB-geo}, with and without the 3.5PN
terms given by $\boldsymbol{\Delta}^{(2)}_{{\sigma}^\ast}$.  As can be
seen, the inclusion of the 3.5 PN terms does not change the ISCO
quantities significantly for $\chi \leq0$, while small differences appear for $\chi > 0$. 
(In the case of $\Omega_{_{\rm ISCO}} M$, however, these differences grow quite 
large when $\chi \rightarrow 1$.) Overall, Figs.~\ref{fig:finalspin}--\ref{fig:omega-isco}
 suggest that the model has reasonable convergence properties for radii $r\geq r_{_{\rm ISCO}}$. 

\begin{figure}[!t]
\includegraphics[width=7cm, clip=true]{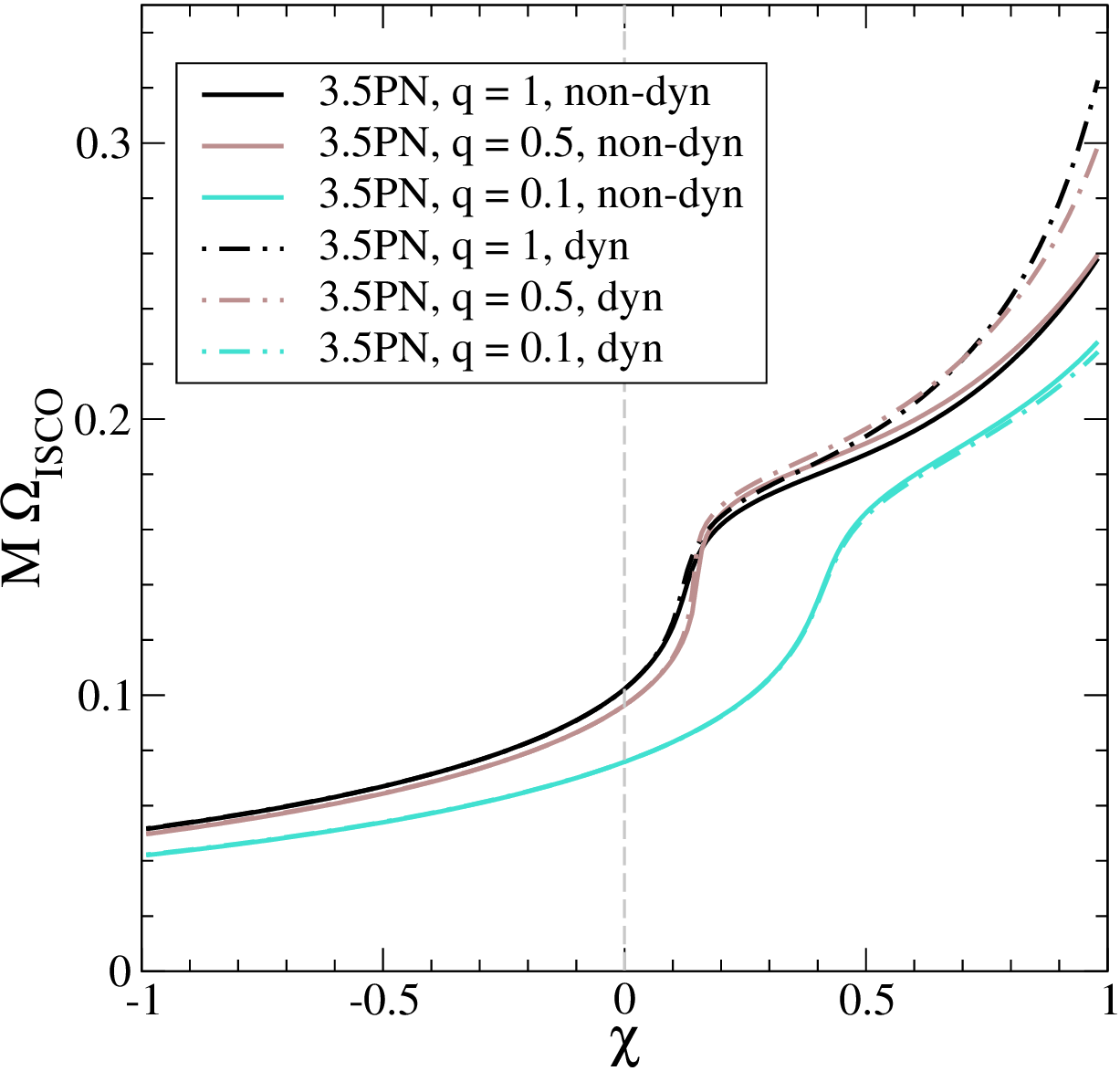}
\caption{\label{omegaISCOCompare} The ISCO frequency for the 3.5PN EOB models with
dynamical (dyn) and non-dynamical (non-dyn) mapping of the spins, for binaries having
spins parallel to $\boldsymbol{L}$, mass ratio $q = m_2/m_1$ and 
spin-parameter projections onto the direction of $\boldsymbol{L}$ given by
$\chi_1 = \chi_2 = \chi$.}
\end{figure}

The results for the model with non-dynamical spin mapping are similar [i.e., a
comparison of the ISCO quantities calculated using the Hamiltonian
\eqref{HeffEOB-ngeo}, with and without the 3.5PN term ${\cal Q}_{\rm
  3.5PN}$, gives similar results]. In general, however, the model with
non-dynamical spin mapping presents lower values for $\Omega_{_{\rm
    ISCO}} M$ at high spins and for comparable mass ratios (see
Fig.~\ref{omegaISCOCompare}, where we compare the 3.5PN models with
dynamical and non-dynamical spin mapping).

\begin{figure}[!t]
\includegraphics[width=7cm, clip=true]{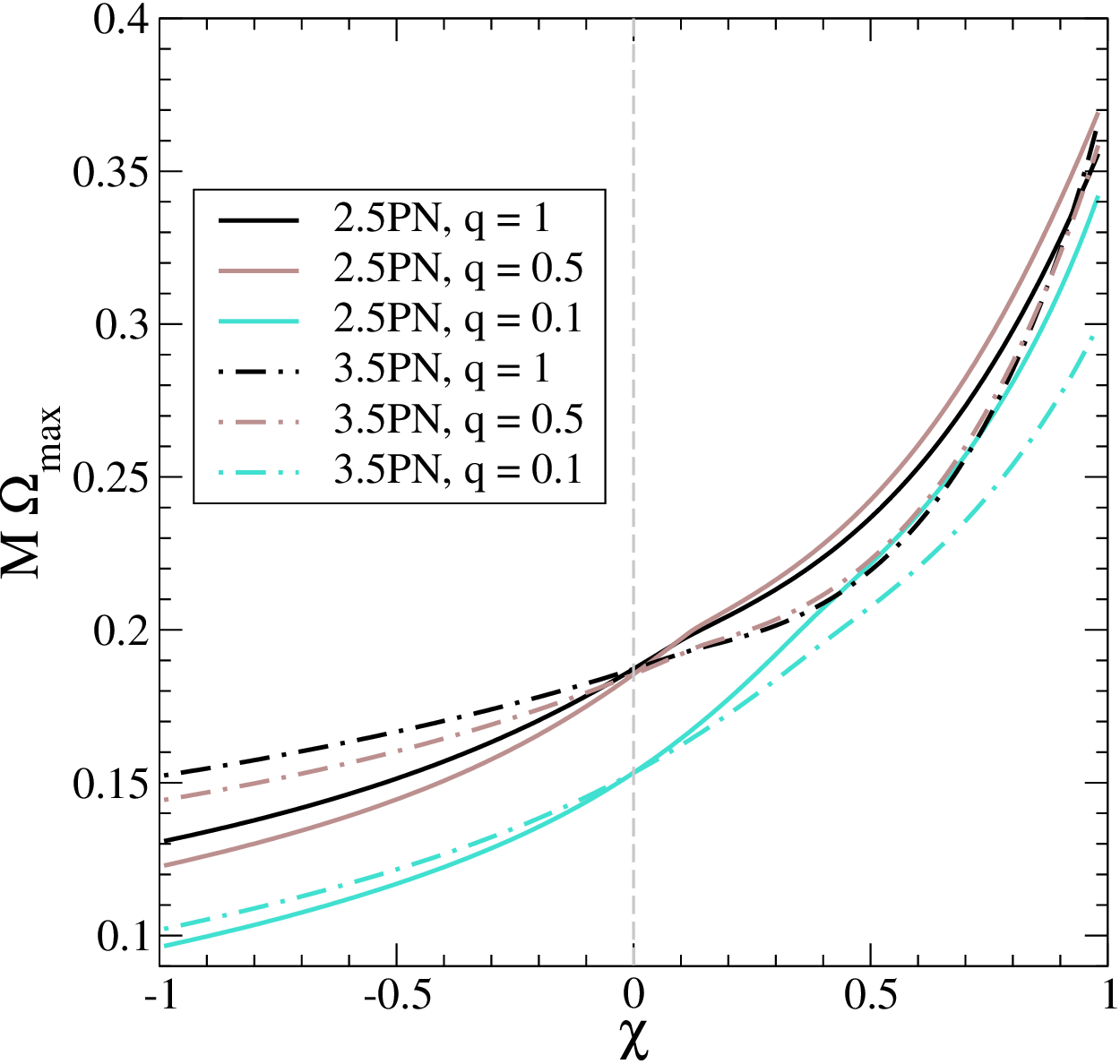}  
\caption{The same as in Fig.~\ref{fig:finalspin}, but for the maximum of the orbital frequency during the plunge.\label{fig:omega-max}}
\end{figure}

\begin{figure}[!t]
\includegraphics[width=7cm, clip=true]{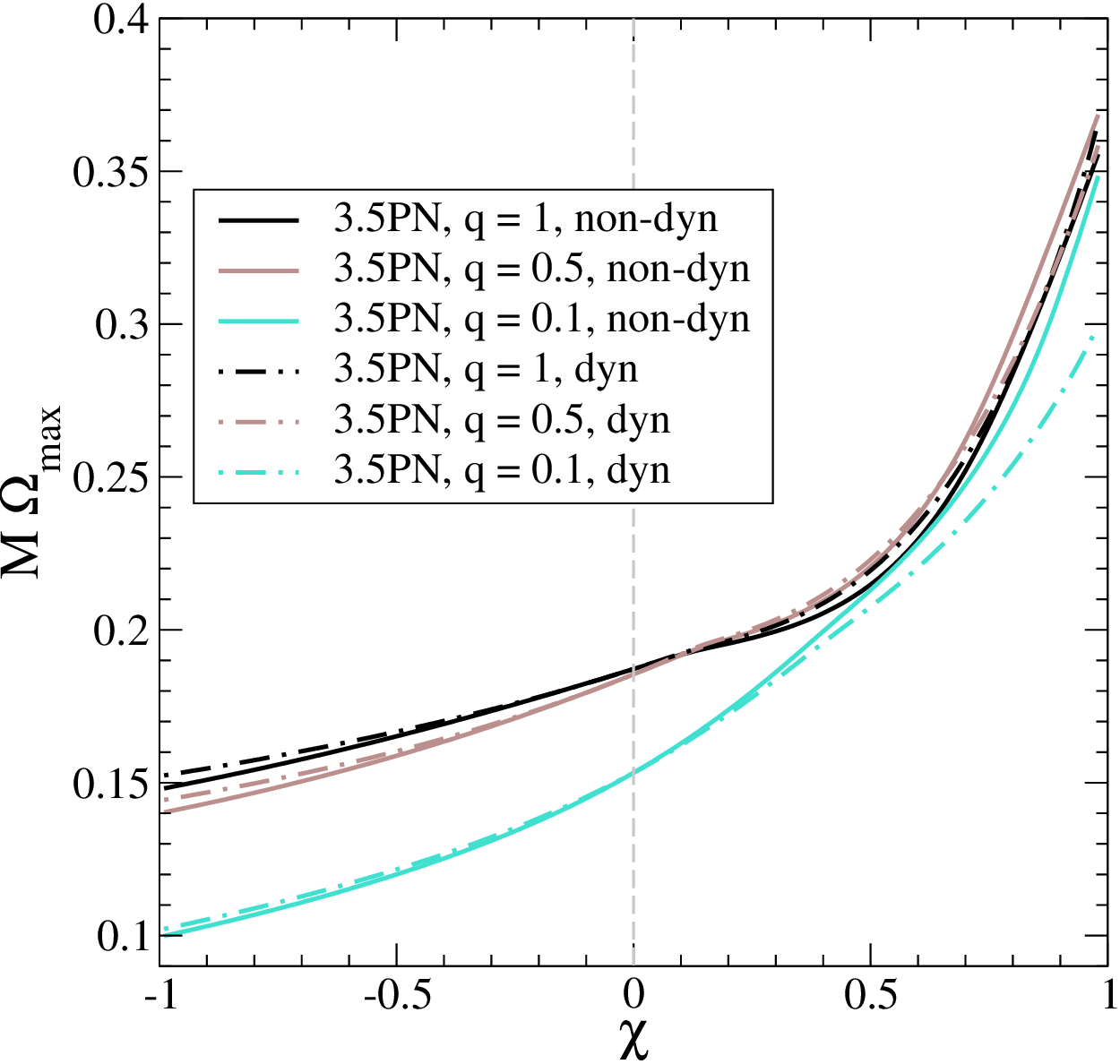}
\caption{\label{omegaMaxCompare} The same as in Fig.~\ref{omegaISCOCompare}, but for the maximum frequency during the plunge.}
\end{figure}

Another attractive feature of our models
is the existence of a peak of the orbital frequency during the plunge starting at the ISCO. More precisely, 
we assume that the effective particle
 starts off with no radial velocity at the ISCO (thus having angular
 momentum $L_{_{\rm ISCO}}$ and energy $E_{_{\rm {ISCO}}}$), and we evolve the geodesic 
equations by calculating the radial momentum
 $p_r$ during the plunge from energy and angular momentum conservation.
 We then calculate the orbital frequency $\Omega=\partial H^{\rm improved}_{\rm real}/\partial L_z$ 
along the trajectory and find that it presents a peak $\Omega_{\max}$. This
is not surprising because the same behavior was observed to be generic in our earlier model~\cite{Barausse:2009xi}. 
The values of $M \Omega_{\max}$ for binaries with spins parallel to $\boldsymbol{L}$, as function of $\chi=\chi_1=\chi_2$,
are shown in Fig.~\ref{fig:omega-max} for mass ratios $q=1, 0.5$ and $0.1$, for the EOB model with dynamical spin mapping at 2.5PN and 3.5PN.
As can be seen the differences introduced by the 3.5 PN terms, although reasonable, are larger than for the ISCO quantities. This may be
because the plunge happens at radii that are smaller than $r_{_{\rm ISCO}}$ and approach the horizon's radius, thus making the higher order
PN terms more and more important. 
The results for the model with non-dynamical spin mapping are generally similar, although
they differ slightly at high spins. In particular, in Fig.~\ref{omegaMaxCompare} we compare the 3.5PN models 
with dynamical and non-dynamical spin mapping.
As can be seen, for $q=0.5$ and $q=1$ the predictions of the two models are very close,
while for $q=0.1$ the model with dynamical spin mapping presents somewhat lower maximum frequencies.

Also, we stress that the values of $M \Omega_{\max}$ for spin antialigned with the angular momentum (i.e., $\chi_1=\chi_2=\chi<0$) 
are quite sensitive to
the values of the gauge parameters $a_0$--$a_3$ and $b_0$--$b_3$. For instance, setting all the gauge parameters to $0$ makes the behavior 
of $M \Omega_{\max}$ with $\chi$ non-monotonic if the 3.5PN models (both with dynamical and non-dynamical spin mapping) 
are considered. This effect does not appear in the 2.5PN models, and can in principle be important for the calibration of 
our model with numerical-relativity simulations.
More details on this will be given in a follow-up paper~\cite{Taracchini:2011}. Even worse, when the 
gauge parameters are set to zero the difference in $M \Omega_{\max}$ between the 2.5PN and 3.5PN models is larger than in Fig.~\ref{fig:omega-max},
a sign that the model probably converges more slowly in this gauge.
In light of this, it seems preferable to use
the gauge parameters \eqref{gaugeparameters_start}--\eqref{gaugeparameters_end}, which by canceling out the radial momentum ${\Delta_r}\,\hat{\boldsymbol{p}}\cdot \boldsymbol{n}/\Sigma$
from $\boldsymbol{\Delta}^{(1)}_{{\sigma}^\ast}$ and
 $\boldsymbol{\Delta}^{(2)}_{{\sigma}^\ast}$ (and from ${\cal Q}_{S\,\rm 2.5PN}$ and ${\cal Q}_{S\,\rm 3.5PN}$) provide
a rather regular and monotonic behavior for $M \Omega_{\max}$ and reasonable differences between the 2.5 and 3.5PN models.

\begin{figure}[!t]
\includegraphics[width=7cm, clip=true]{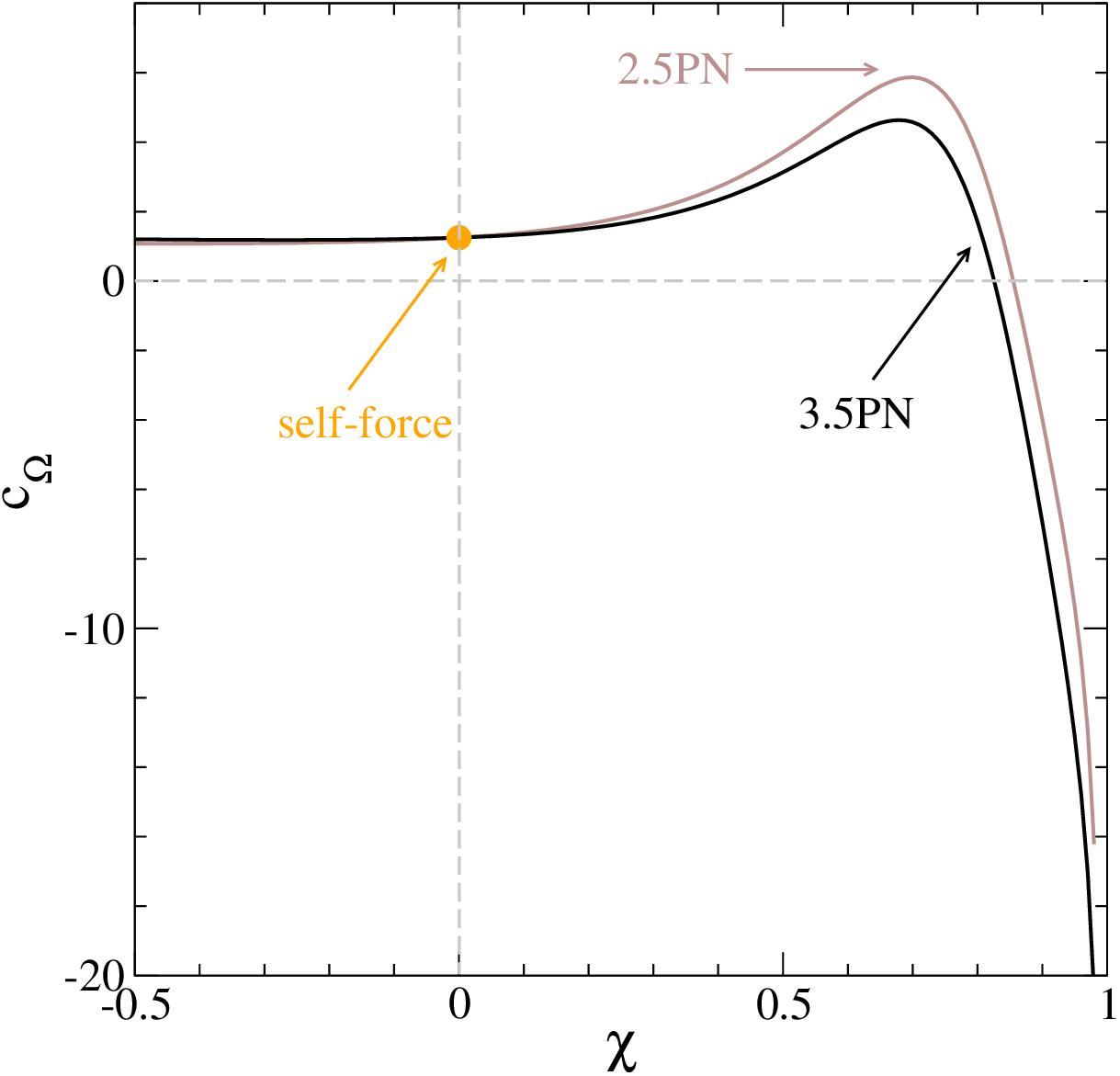} 
\caption{The shift of the ISCO frequency $c_{\Omega}$, defined in Eq.~\eqref{comega}, for the 2.5PN and 3.5PN EOB models with
dynamical mapping of the spins, for a binary having
spins parallel to $\boldsymbol{L}$, mass ratio $q = m_2/m_1=10^{-3}$ and 
spin-parameter projections onto the direction of $\boldsymbol{L}$ given by
$\chi_1 = \chi$ and $\chi_2 =0$. \label{fig:omega-isco-shift}}
\end{figure}

Finally, in Fig.~\ref{fig:omega-isco-shift} we show the predictions of our EOB model with dynamical spin mapping for the ISCO frequency of
a system with $q=m_2/m_1=10^{-3}$, $\chi_1=\chi$ and $\chi_2=0$ (the results for the EOB model with non-dynamical 
spin mapping are similar). More precisely, we show the fractional deviation
from the Kerr ISCO frequency normalized by the mass ratio,
\begin{equation}\label{comega}
c_{\Omega}=\frac{1}{q}\left(\frac{\Omega_{_{\rm ISCO}} M\vert_{q}}{\Omega_{_{\rm ISCO}} M\vert_{\rm Kerr}}-1\right)\,,
\end{equation}
as a function of $\chi$, as proposed in Ref.~\cite{favata}. 
This ISCO shift is caused by the conservative part of the self-force and 
has been calculated exactly by Ref.~\cite{BarackSago09} in the case of a Schwarzschild spacetime ($\chi=0$).
The results of Ref.~\cite{BarackSago09} is $c_{\Omega}=1.2513+{\cal O}(q)$ [see also Ref.~\cite{Damour:2009sm}], 
and is denoted by a filled circle in Fig.~\ref{fig:omega-isco-shift}. As can be seen, both the 2.5 and 3.5PN models predict $c_\Omega>0$, except when
$\chi\gtrsim 0.83$. This change of behavior of the EOB prediction is common also to our earlier model of Ref.~\cite{Barausse:2009xi},
and might have important implication for configurations that might violate  
the Cosmic Censorship Conjecture~\cite{CCC1,CCC2}. However, the behavior of $c_\Omega$,
which seems to diverge as $\chi$ approaches 1, suggests that this might simply be a spurious
effect due to the incomplete knowledge of the  function $K$ [Eq.~\eqref{Keta}] and to the fact that the EOB model
only reproduces the SS coupling at leading PN order (2PN).
As mentioned in Ref.~\cite{Barausse:2009xi}, $K$ may in general depend not only on $\eta$ but also on $\chi^2$, and these spin-dependent
terms can be very important for near-extremal spins, and so will the 3PN SS couplings.

It is therefore possible that after reconstructing the full functional
form of $K$ (by comparing to future self-force calculations in Kerr or 
to numerical-relativity simulations for spinning binaries) and extending the EOB model to include the 3PN SS couplings, $c_\Omega$
might remain positive even at high spins.

\section{Conclusions}
\label{sec:concl}
Recently, Ref.~\cite{Hartung:2011te} has computed the 3.5PN SO effects in the ADM Hamiltonian. We have 
taken advantage of this result and extended the EOB Hamiltonian of spinning black holes to 
include these higher-order SO couplings. 

Building on previous work~\cite{Damour01c,Damour:2007nc}, and in particular on the EOB Hamiltonian
of Refs.~\cite{Barausse:2009aa,Barausse:2009xi}, which 
reproduces the SO test-particle couplings exactly at all PN orders, 
we have worked out two classes of EOB Hamiltonians, 
which differ by the way the spin variables are mapped between the effective and real descriptions. One class 
of EOB Hamiltonians is the straightforward extension to the next PN order of the EOB Hamiltonian of 
Ref.~\cite{Barausse:2009xi}. It uses a mapping between the real and effective spin variables that depends 
on the dynamical orbital variables $\mathbf{p}^2$, $\mathbf{n} \cdot \mathbf{p}$  and $r$. By contrast, the other class 
of EOB Hamiltonians uses a mapping between the real and effective spin variables that {\it does not} depend 
on these dynamical orbital variables. We achieved this result at the cost of modifying the Hamilton-Jacobi equation of a spinning test-particle. 

Quite interestingly, when restricting to spins aligned or antialigned with the orbital 
angular-momentum and to equatorial circular orbits, we find that the predictions of 
these two classes of EOB Hamiltonians for the 
ISCO frequency, energy and angular momentum, and for the maximum of the orbital frequency 
during the plunge are generally similar. However, for high spins
the model with dynamical mapping of the spins
 may present somewhat lower maximum frequencies and larger ISCO frequencies.

As pointed out originally in Ref.~\cite{Damour:2007nc}, several gauge parameters can 
enter the canonical transformation that maps the real and effective Hamiltonians. 
If the Hamiltonian were known exactly, i.e., at all PN orders, then 
physical effects should not depend on these parameters. However, since we know the 
Hamiltonian only at a certain PN order, we expect these gauge parameters 
to lead to non-negligible differences. In fact, we obtained that when setting all the gauge 
parameters to zero, the maximum frequency during the plunge has a non-monotonic dependence on the
spins, and varies quite significantly as a consequence of the inclusion of the 3.5 PN SO couplings. We found instead that
when choosing the gauge parameters so that the terms depending on the radial momentum disappear from our spin mapping (in the model
with dynamical spin mapping) or from the modifications to the Hamilton-Jacobi equation (in the  model
with non-dynamical spin mapping), the maximum frequency during the plunge has a much more regular behavior and varies by small amounts
when adding the 3.5PN SO couplings. This suggests that such a choice of the gauge parameters may accelerate the convergence
of the model's results in the strong-field region where the plunge takes place.

The EOB Hamiltonians derived in this paper can be calibrated to numerical-relativity 
simulations with the goal of building analytical templates for LIGO and Virgo searches. 
A first example was obtained in Ref.~\cite{Pan:2009wj}, where the EOB Hamiltonian at 2.5PN 
order in the SO couplings of Ref.~\cite{Damour:2007nc} was calibrated to two highly-accurate 
numerical simulations. Results that use the EOB Hamiltonian at 3.5PN order 
developed in this paper will be reported in the near future~\cite{Taracchini:2011}.

Lastly, while finalizing this work,  Ref.~\cite{2011arXiv1106.4349N} appeared 
in the archives as a preprint. Both this paper and Ref.~\cite{2011arXiv1106.4349N} derive the 
effective gyromagnetic coefficients [see Eq.~(\ref{gyroeffSO})], but with two different methods. 
Our computation uses the Lie method to generate both the purely-orbital and 
the spin-dependent canonical transformations, while  Ref.~\cite{2011arXiv1106.4349N} 
first applies explicitly the  purely-orbital transformation from ADM to EOB coordinates, and then uses Eq.~\eqref{Htransf}
to account for the effect of a spin-dependent canonical transformation. 
As a result of these different procedures, and as discussed in Sec.~\ref{sec:ADMH}, the 
2.5PN gauge parameters in our spin-dependent canonical transformation coincide 
with those of Ref.~\cite{2011arXiv1106.4349N}, but the 3.5PN gauge parameters have different meanings
in the two approaches and therefore do not coincide.
However, by suitably expressing our 3.5PN gauge parameters 
in terms of those of Ref.~\cite{2011arXiv1106.4349N}, we find that 
our effective gyromagnetic coefficients fully agree with those
of Ref.~\cite{2011arXiv1106.4349N}. This amounts to saying that our gyromagnetic coefficients agree 
with those of Ref.~\cite{2011arXiv1106.4349N} up to a canonical transformation, and are therefore
physically equivalent. 
 More importantly, in this paper we have focused on and worked out two classes 
of EOB Hamiltonians that are different from the one considered in Ref.~\cite{2011arXiv1106.4349N}.

\begin{acknowledgments} 
E.B. and A.B. acknowledge support from NSF Grant PHY-0903631.
A.B. also acknowledges support from NASA grant NNX09AI81G. 
\end{acknowledgments} 

\bibliography{References}

\end{document}